\documentclass{article}
\usepackage{graphicx} % Required for inserting images
\usepackage{slashed}
\usepackage{amsmath}
\usepackage{authblk}
\usepackage{comment}
\usepackage{hyperref}
\usepackage{multirow}
\usepackage{cite}

\title{Portals to New Physics: The W-Quark Portal}
\author[1]{Linda M. Carpenter\thanks{lmc@physics.osu.edu}}
\author[1]{Katherine Schwind\thanks{schwind.44@osu.edu}}
\author[2]{Taylor Murphy\thanks{murphyt6@miamioh.edu}}

\affil[1]{\emph{Department of Physics, The Ohio State University,
191 W. Woodruff Avenue, Columbus, OH 43210, U.S.A.}}
\affil[2]{\emph{Department of Physics, Miami University, 500 E. Spring St., Oxford, OH 45056, U.S.A.}}

\begin{document}

\maketitle
\begin{abstract}
We explore new EFT operators where single new exotic states interact with the
Standard Model through an asymmetric Standard Model portal with couplings
to at least one quark and one W boson. Effective operators up to dimension 6 are considered, and we specify SM charges of all new states accessible through this portal. We explore our interactions by detailing possible collider production processes
at e-p and p-p colliders. Finally, we study the LHC production of a new weak-quadruplet SU(3) fundamental fermion which can be singly produced in association with a SM quark through W-quark portal processes.
\end{abstract}

\newpage
\section{Introduction}

In this work, we investigate all possible Beyond the Standard Model (BSM) states that one can access through new effective operators, up to dimension 6, that are part of the W-quark portal. These operators include at least one quark and one W boson, along with one exotic BSM state. Other SM states are allowed, as well. We have begun exploring portal-based approaches to categorizing new BSM physics interactions with our recent work, and this paper is a continuation of this effort \cite{Carpenter:2024hvp,murphy2024quarkleptonportalleptoquarks,Carpenter:2021gpl}. Specifically, we have examined new interactions between beyond ``Light Exotic" (LEX) particles and the
particles of the SM \cite{Carpenter:2023giu}. We use a light exotics effective field
theory (LEX-EFT) to categorize all new species of exotic particle that can interact
with the Standard Model through a specified set of SM fields. We capture these interactions with a set of effective
operators that specify interactions between particles. These interactions are characterized
by an effective mass cutoff scale $\Lambda$.

\begin{equation}
L \supset \frac{1}{\Lambda^n}\Psi_i...\Psi_n\phi_{\alpha}\phi_{\beta}... 
\end{equation}

Here, the fields $\Psi$ are exotic states, the fields $\phi$ are Standard
Model states, and $\Lambda$ is the scale of the effective cutoff.
Adding such interaction terms to the SM Lagrangian captures many phenomenologically useful features. From these terms we are able to obtain information on kinematics, faithful prediction of particle scattering
amplitudes up to the validity threshold of the effective field theory,
and a picture of the charge flow through the interaction. The above
operators must be singlets under all Standard Model and Beyond the
Standard Model symmetries.

The motivation behind this approach lies in the problem that we do not know what particles may exist
Beyond the Standard Model. In fact, only a small fraction of all possible
models have been thoroughly explored. However, many unexplored models have extremely novel
and non-standard phenomenology. These may present truly new discovery channels
at current and future particle colliders. 

In the portal-based effective field theory (EFT) approach, we take a systematic
view of categorizing possible exotic models and the particle phenomenology
that results. The approach involves fixing a set of Standard Model
states in effective operators. These portals then determine specific sets
of new particle production processes at particle colliders. We can
thus categorize new classes of LEX states that are accessible through
each production portal, as well as specify all of the relevant event
topologies involved in the production and decay of these particles.
Specific portals are useful for determining which new classes of exotically charged particles can be reached through interactions with SM particles, and for specifying the exact regions of exotic
theory space that can be reached through specific SM processes.

Here, we continue the trend of investigation into a single portal, and focus on the W-quark portal.
In the case of this work, we intend to only study operators that would
lead to the single production and decay of a light exotic state. This means
that each LEX-EFT operator under consideration will only contain a single
LEX BSM state. We note that LEX fields that are charged
under SM gauge groups can always be pair-produced at particle colliders
by SM processes. However, the single production of exotic states
often serves as a more powerful discovery channel \cite{Carpenter:2024qti}. For new particles
in the TeV mass range, single production of heavy states remains possible
in mass ranges where pair production becomes kinematically inaccessible.
In addition, single production generally produces cleaner collider
events, making it more likely to carry higher search efficiencies. 

The W-quark portal, in particular, allows us to access new exotic
states that are in highly non-trivial representations of both SU(3)
and SU(2). While there have been previous phenomenological studies
of exotics in triplet, sextet and octet representations of SU(3) \cite{Carpenter:2021rkl,Carpenter:2022qsw,Carpenter:2023aec,Chen:2008hh,Cao:2013wqa,Carpenter:2021vga,Carpenter:2015gua}, interactions are usually assumed to be renormalizable and exotic colored states are
generally assumed to be singlets of SU(2). Some more expansive exotics studies have generally focused on di-quark signatures of exotic colored scalars \cite{Blum:2016szr,Han:2010rf,Han:2023djl,Manohar:2006ga,Miralles:2019uzg,Carpenter:2011yj,Celikel:1998dj}. These studies have begun to open up considerations to more general particle species and/or interactions. 

As we will see, the W-quark
portal allows us to discover even more general interactions for many novel species of exotics, and it will allow special access to LEX states of high-dimensional
representations of SU(2) and SU(3). These states have particularly interesting cascade
features. In addition, we will consider various hadron collider
processes for the single production of exotic states. The W-quark portal
allows us to access single-exotic production processes
that involve a hard initial-state quark. One feature of the LHC is
that, at an intermediate momentum fraction, incoming gluons have the highest
parton distribution functions. However, at a truly large momentum fraction,
quarks have the largest parton distribution functions \cite{ParticleDataGroup:2024cfk}. Thus, the W-quark
portal may have a reach to produce significantly more massive exotic states than those of the W-gluon portal, which we considered in a previous work \cite{Carpenter:2024hvp}.

We also note that in our models, we are not assuming any particular
BSM symmetries. In fact, such symmetries would serve to restrict the
number of effective operators coupling a LEX state to the SM. In this
sense, the catalog we write in this paper is a maximal set in which
the addition of various BSM symmetries restricts us to various subsets.

We will now go on to specify all possible EFT interactions up to dimension
6 that a single light exotic state can have with the W-quark portal.
Each of the operators written below should allow interactions between at least one quark,
one W boson, and one LEX state. As shown in Fig. 1, if we consider $X$ to
be our LEX state, these interactions involve a single particle $X$, one quark,
and one W boson, either alone or with other SM particles.

\begin{center}
\includegraphics[scale=.13]{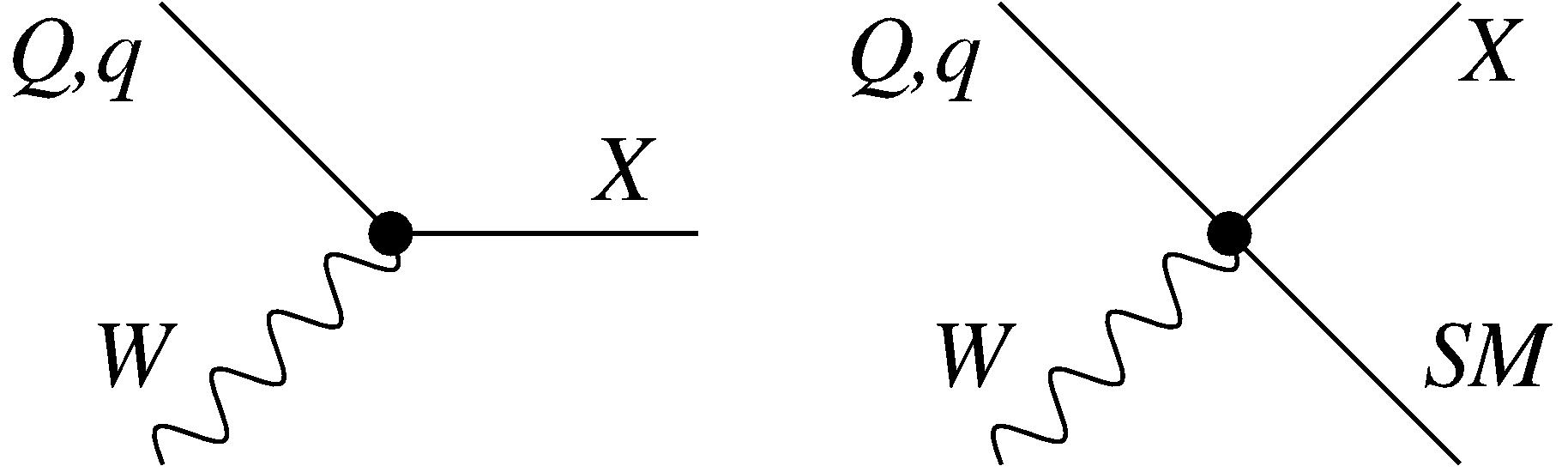} \\
Figure 1. Possible interaction vertices in the W-quark portal
\end{center}

We will catalog the various interactions and specify the maximal set
of allowed quantum numbers a LEX state may take in these interactions.
We will then categorize the allowed collider production processes of
our LEX states for these interactions.

A final note is that, in enumerating a complete set of effective interactions
of the W-quark portal, we find a theoretical degeneracy between operators. As pointed out in studies of SMEFT,
integration by parts, various quantum field theory identities, or
lower-order equations of motion may relate effective operators \cite{Grzadkowski_2010,Murphy:2020rsh,Lehman:2014jma}. In
our previous work, we specified that, in defining our portal, we remove
any redundancy of operators involved in integration by parts or various
QFT identities. This, along with the equations of motion, forces us to
choose a basis for our EFT. We recognize that through change of basis,
some of the interactions we list can be moved from our portal into
a different phenomenological portal; however, the physical production
processes and measurable phenomenology will be exactly the same. Since
we want our work to be of maximum utility in categorizing the phenomenology
of new models, we choose phenomenological bases that maximize the
number of operators we consider in the portal. 

This work proceeds as follows. In Section 2, we discuss operators without derivatives which contain spin-0 (scalar) LEX states. Section 3 enumerates operators without derivatives which contain spin-1/2 (fermion) LEX states. Sections 4 and 5 again discuss operators including scalar and fermion LEX states, respectively. In these sections, the operators also include at least one covariant derivative. Finally, Section 6 demonstrates some collider phenomenology in the W-q portal by exploring LHC collider production for one of the new exotic states in the W-q portal, the fermionic SU(2) quadruplet color triplet. Section 7 concludes.

\section{Spin 0 LEX states}

We will now enumerate the operators of the W-quark portal up to effective
dimension 6. These operators will all contain at least one SM quark, one W boson and one BSM light exotic (LEX) state. They will thus constitute the set of all operators through which an exotic LEX state may be reached though W-quark interactions.  The operators must be gauge singlets and are  found  following
an iterative tensor product method. For an N particle operator, the possible representation of the N-1 SM field products are found by nested application of  the 2-field tensor product rules. The quantum numbers of the LEX state are then chosen to ensure a singlet. In the tables below, we will specify the form of the operator,
in addition to its dimension and SU(3), SU(2), and U(1)$_Y$ quantum numbers of the LEX state involved.

In this section we consider the LEX state to be a spin 0 CP even scalar and we  will enumerate operators that do not contain  derivatives. 
In each of the operators below, the scalar LEX field is denoted by $\phi$. SU(2) indices are written in lowercase Roman letters, and Lorentz indices are indicated with Greek letters. We choose not to explicitly write out SU(3) indices.

\subsection{Operators with $\boldsymbol{Qf\phi W}$}

In the case that the LEX field is a scalar and the operator contains no derivatives, the operator must contain a quark, the LEX state, and a W boson which comes from the electroweak field strength tensor $W_{\mu\nu}$. All operators must contain one other SM fermion. The second SM fermion can be either a quark or a lepton. Each of the SM fermion fields can be left-handed or right-handed. By necessity, these operators will be dimension 6. 

We start by listing all operators of this type which include only right-handed SM fermions. Operators including only one quark will have a LEX state with an SU(3) quantum number of 3. Operators with two quarks can take two different forms. For operators with one quark and one anti-quark, the SU(3) LEX quantum number follows from the SU(3) tensor product rule \cite{Coleman:1965afp,Slansky:1981yr}
\begin{equation}
3\otimes\overline{3}= 1+8 \ .
\end{equation}
As such, the LEX state may be  a singlet or octet. Conversely, for operators with a quark and a charge conjugate of an anti-quark, the SU(3) LEX representation follows from the SU(3) tensor product rule
\begin{equation}
3\otimes 3 = 3+6 \ .
\end{equation}
Thus, the LEX state must be an SU(3) triplet or sextet.

For the operators of Table 1, which contain no left-handed SM fermions, the SU(2) quantum number of the LEX state must contract directly with the SU(2) field strength tensor. As such, these LEX states will be SU(2) triplets.
\begin{center}
\begin{tabular}{|c|c|c|}
\hline 
dimension & Operators with scalar LEX fields & (SU(3),SU(2),Y)\tabularnewline
\hline 
\hline 
dim 6 & $\overline{u^c} \sigma^{\mu \nu} \ell \phi_{i j} W_{\mu \nu}^{i j}  $ & ($\overline{3},3,1/3$)\tabularnewline
\hline 
dim 6 & $\overline{d^c} \sigma^{\mu \nu} \ell \phi_{i j} W_{\mu \nu}^{i j}  $ & ($\overline{3},3,4/3$)\tabularnewline
\hline 
dim 6 & $\overline{u^c} \sigma^{\mu \nu} u \phi_{i j} W_{\mu \nu}^{i j}  $ & ($\overline{6},3,-4/3$) , ($3,3,-4/3$)\tabularnewline
\hline 
dim 6 & $\overline{u^c} \sigma^{\mu \nu} d \phi_{i j} W_{\mu \nu}^{i j}  $ & ($\overline{6},3,-1/3$) , ($3,3,-1/3$)\tabularnewline
\hline 
dim 6 & $\overline{d^c} \sigma^{\mu \nu} d \phi_{i j} W_{\mu \nu}^{i j}  $ & ($\overline{6},3,2/3$) , ($3,3,2/3$)\tabularnewline
\hline 
\end{tabular}
\\
Table 1. Dimension 6 operators containing $Qf\phi W$ with only right-handed SM fermions
\end{center}
Each operator above contains one quark and one charge conjugate anti-quark; thus, LEX states are color triplets or sextets.

We note that the first two operators in Table 1 contain one quark and one lepton. In order to conserve some discreet SM charges, we must assign the LEX state both lepton and baryon number. These lepton and baryon number assignments are the same as those of scalar lepto-quarks \cite{BUCHMULLER1987442}. Indeed, these states are extended SU(2) representations of scalar lepto-quarks. Some of these lepto-quark SU(2) adjoints have been studied in the context of possible di-quark couplings \cite{Blum:2016szr}. However, the effective operators above are novel dimension 6 interactions that have not previously been studied. The bottom three operators involve a quark and charge conjugate anti-quark; to conserve baryon number the LEX states must be assigned a double baryon number of $-2/3$. Thus the W-quark portal leads to novel color sextet and and triplet scalars with double baryon number.

Next, we examine operators of this type which contain one left-handed SM fermion and one right-handed SM fermion. The SU(2) gauge fields are in the adjoint representation of SU(2). With one weak doublet, we can look to the SU(2) product rule
\begin{equation}
 3 \otimes 2 = 2 \oplus 4   
\end{equation}
to see that the LEX state must be a doublet or quadruplet under SU(2). These operators are found in Table 2.

\begin{center}
\begin{tabular}{|c|c|c|}
\hline 
dimension & Operators with scalar LEX fields & (SU(3),SU(2),Y)\tabularnewline
\hline 
\hline 
dim 6 & $\overline{Q_{L}}_{i} \sigma^{\mu \nu} \ell \phi_j W_{\mu \nu}^{i j}  $ & ($3,2,7/6$)\tabularnewline
\hline 
dim 6 & $\overline{Q_{L}}^{i} \sigma^{\mu \nu} \ell \phi_{i j k} W^{j k}_{\mu \nu}   $ & ($3,4,7/6$)\tabularnewline
\hline 
dim 6 & $\overline{Q_{L}}_{i} \sigma^{\mu \nu} u \phi_j W_{\mu \nu}^{i j}$ & ($1,2,-1/2$) , ($8,2,-1/2$)\tabularnewline
\hline
dim 6 & $\overline{Q_{L}}^{i} \sigma^{\mu \nu} u \phi_{i j k} W_{\mu \nu}^{j k}$ & ($1,4,-1/2$) , ($8,4,-1/2$)\tabularnewline
\hline
dim 6 & $\overline{Q_{L}}_{i} \sigma^{\mu \nu} d \phi_j W_{\mu \nu}^{i j}$ & ($1,2,1/2$) , ($8,2,1/2$)\tabularnewline
\hline
dim 6 & $\overline{Q_{L}}^{i} \sigma^{\mu \nu} d \phi_{i j k} W_{\mu \nu}^{j k}$ & ($1,4,1/2$) , ($8,4,1/2$)\tabularnewline
\hline
dim 6 & $\overline{u} \sigma^{\mu \nu} L_i \phi_j W_{\mu \nu}^{i j}  $ & ($3,2,7/6$)\tabularnewline
\hline 
dim 6 & $\overline{u} \sigma^{\mu \nu} L^i \phi_{i j k} W_{\mu \nu}^{j k}  $ & ($3,4,7/6$)\tabularnewline
\hline 
dim 6 & $\overline{d} \sigma^{\mu \nu} L_i \phi_j W_{\mu \nu}^{i j}  $ & ($3,2,1/6$)\tabularnewline
\hline 
dim 6 & $\overline{d} \sigma^{\mu \nu} L^i \phi_{i j k} W_{\mu \nu}^{j k}  $ & ($3,4,1/6$)\tabularnewline
\hline 
\end{tabular}
\\
Table 2. Dimension 6 operators containing $Qf\phi W$ with one left-handed SM fermion
\end{center}

In Table 2, operators containing a quark and an anti-quark yield a LEX state in a singlet or octet representation of SU(3), while operators containing one quark and one lepton involve LEX states that are SU(3) triplets. Again, we note that operators in Table 2 which contain one lepton and one quark require the LEX state to have both lepton number $-1$ and baryon number $1/3$; the LEX states in these operators are SU(3) triplets and thus are types of scalar lepto-quark. Operators with a quark and anti-quark require the LEX state to have baryon number 0. These are the exotic SU(3) singlets and octets. The particle in the ($1,2,-1/2$) representation has the quantum numbers of an exotic Higgs field, while the state in the ($8,2,-1/2$) representation has quantum numbers of the exotic Manohar-Wise octet \cite{Manohar:2006ga}. The W-quark portal also allows us access to exotic SU(3) singlet and octets which are 4's of SU(2) and thus contain doubly charged states. 

Finally, we consider operators of type $Qf\phi W$ that contain two left-handed SM fermions. These operators contain the SU(2) triplet gauge field and two doublet fermions. The SU(2) tensor products of SM fields yield
\begin{equation}
3 \otimes 2 \otimes 2 = (2 \oplus 4) \otimes 2 \rightarrow 1 \oplus 3 \oplus 5
\end{equation}
As such, the LEX state can have a singlet, triplet, or quintuplet representation under SU(2). Operators containing one quark and one lepton will involve SU(3) triplet LEX states while those containing 2 SM quarks involve SU(3) triplet and sextet LEX states.  

\begin{center}
\begin{tabular}{|c|c|c|}
\hline 
dimension & Operators with scalar LEX fields & (SU(3),SU(2),Y)\tabularnewline
\hline 
\hline 
dim 6 & $\overline{Q_{L}^c}_{i} \sigma^{\mu \nu} L_j \phi W_{\mu \nu}^{i j}$ & ($\overline{3},1,1/3$)\tabularnewline
\hline
\multirow{2}{*}{dim 6} & $\overline{Q_{L}^c}_{i} \sigma^{\mu \nu} L_j \phi^j_k W_{\mu \nu}^{i k}$ & \multirow{2}{*}{($\overline{3},3,1/3$)}\tabularnewline
 & $\overline{Q_{L}^c}_{i} \sigma^{\mu \nu} L_j \phi^i_k W_{\mu \nu}^{j k}$ & \tabularnewline
\hline
dim 6 & $\overline{Q_{L}^c}^{i} \sigma^{\mu \nu} L^j \phi_{i j k l} W_{\mu \nu}^{k l}$ & ($\overline{3},5,1/3$)\tabularnewline
\hline
dim 6 & $\overline{Q_{L}^c}_{i} \sigma^{\mu \nu} Q_{L j} \phi W_{\mu \nu}^{i j}$ & ($\overline{6},1,-1/3$) , ($3,1,-1/3$)\tabularnewline
\hline
\multirow{2}{*}{dim 6} & $\overline{Q_{L}^c}_{i} \sigma^{\mu \nu} Q_{L j} \phi^i_k W_{\mu \nu}^{j k}$ & \multirow{2}{*}{($\overline{6},3,-1/3$) , ($3,3,-1/3$)}\tabularnewline
 & $\overline{Q_{L}^c}_{i} \sigma^{\mu \nu} Q_{L j} \phi^j_k W_{\mu \nu}^{i k}$ & \tabularnewline
\hline
dim 6 & $\overline{Q_{L}^c}^{i} \sigma^{\mu \nu} Q_L^j \phi_{i j k l} W_{\mu \nu}^{k l}$ & ($\overline{6},5,-1/3$) , ($3,5,-1/3$)\tabularnewline
\hline
\end{tabular}
\\
Table 3. Dimension 6 operators containing $Qf\phi W$ with only left-handed SM fermions
\end{center}

One again, operators that contain one quark and one lepton give leptoquark-like discreet charge assignment to the LEX state. Operators that involve one quark and one charge conjugated anti-quark assign the LEX state a double baryon number of $-2/3$. This assignment, along with the unusual SU(2) and SU(3) LEX charge assignments, shows how the W-quark portal can access very unusual theory space.

\subsection{Some Collider Phenomenology}
From the above lists, which contain operators with a scalar LEX field and no derivatives, we can identify several interesting LEX states. In particular, we see states that are fundamentals, sextets, and octets under SU(3) that also have a significant SU(2) charge. Specifically, we have several instances of triplet, quadruplet, and even quintuplet SU(2) states.  We also see states that carry baryon number $-2/3$ and other states with unit lepton number and baryon number $1/3$. These latter states can even carry the same quantum numbers as standard leptoquarks, resulting in new effective interactions for standard lepto-quark theories. We discussed this idea a bit in explorations of the lepton-quark portal \cite{murphy2024quarkleptonportalleptoquarks}.

We now choose to examine one of our fields from Section 2 that has the highest SU(2) number available, along with an octet SU(3) representation. This field, with quantum numbers $(8,5,-1/3)$, can be found in the dimension 6 operator $\frac{1}{\Lambda^2}\overline{Q_{L}^c}^{i} \sigma^{\mu \nu} Q_L^j \phi_{i j k l} W_{\mu \nu}^{k l}$. The scalar field in this operator has electrically charged components ($5/3, 2/3, -1/3, -4/3, -7/3$). These multiply-charged components are quite interesting.

From operators in this section, we find a few different types of 4-particle and 5-particle vertices. The four particle interactions, involving a single gauge boson, are depicted in Fig. 2. 
\begin{center}
\includegraphics[scale=.12]{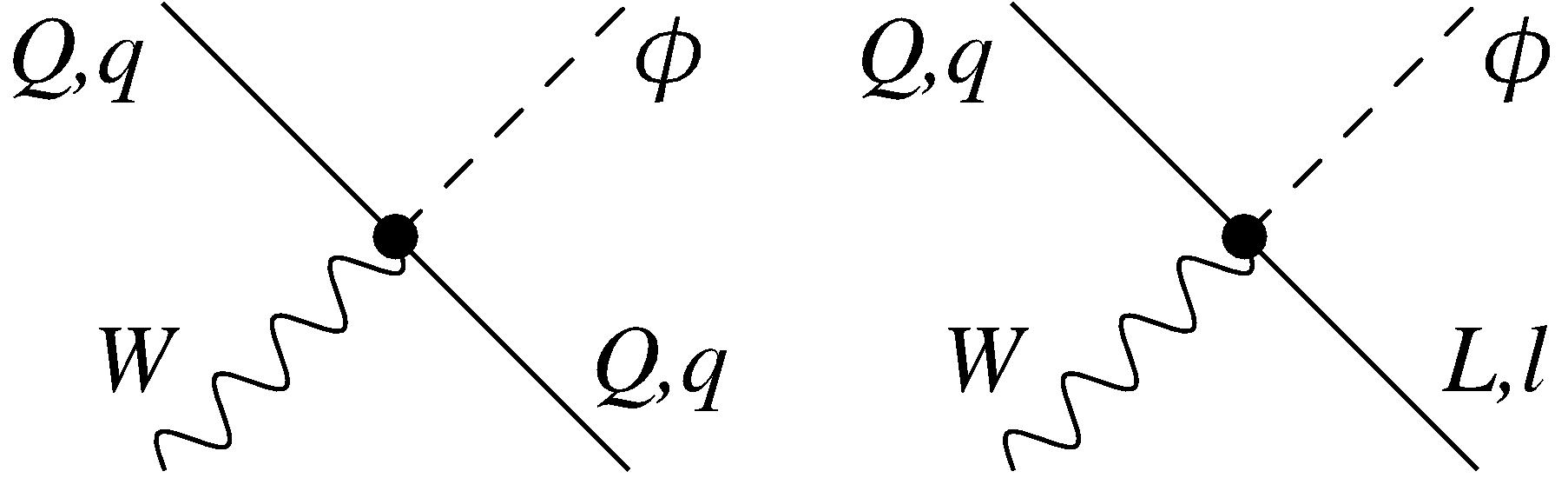} \\
    Figure 2. Four-point vertices involving a scalar LEX state in the W-quark portal
\end{center}
The SU(2) field strength tensor can also result in terms with two gauge bosons. This results in 5-point interactions, as shown in Fig. 3.

\begin{center}
\includegraphics[scale=.1]{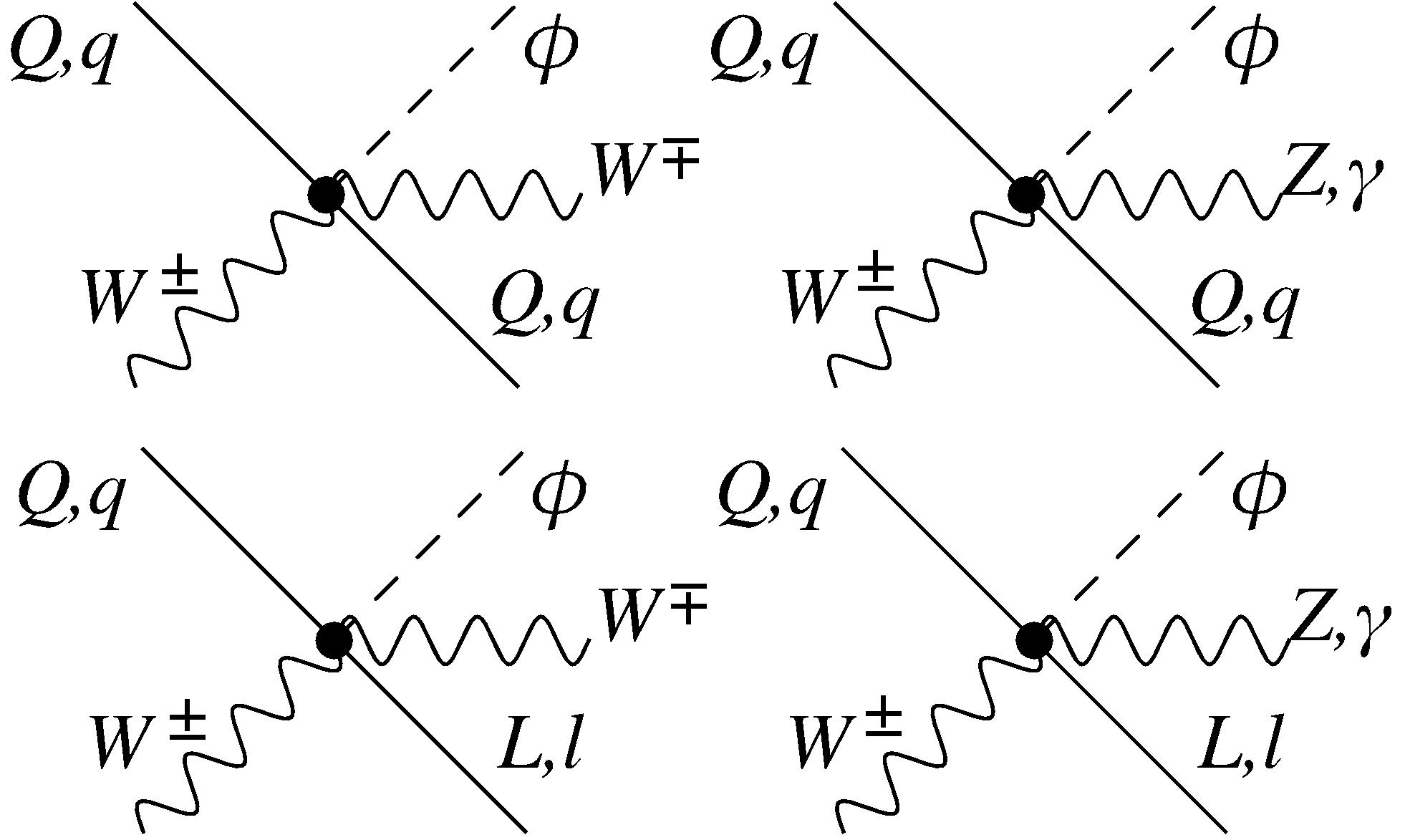} \\
    Figure 3. Five-point vertices involving a scalar LEX state in the W-quark portal
\end{center}

A variety of processes can occur at colliders due to these types of interactions. In Fig. 4, we show two possible associated production processes at two different types of collider. In the left panel of Fig. 4, we show an example process which would occur at proton-proton colliders, such as the LHC. In this process, an off-shell W boson collides with quark to produce a LEX scalar and quark. The initial-state quark which radiated the W boson is also present in the final state particles. This process is $qq \rightarrow q q \phi$. 

\begin{center}
    \includegraphics[scale=.15]{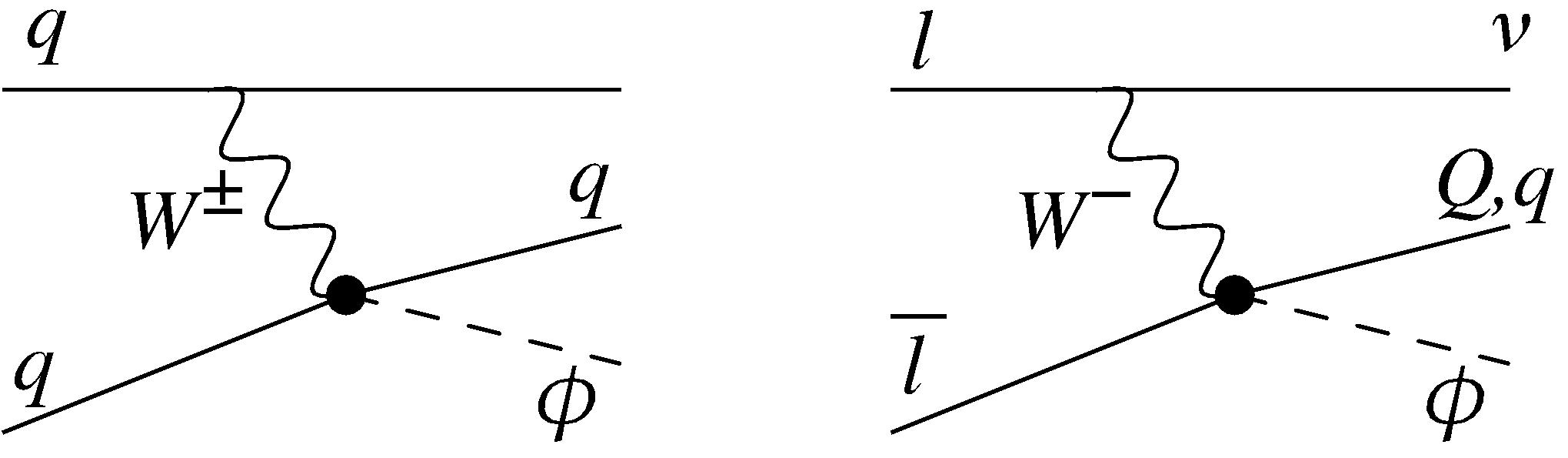}\\
    Figure 4. Associated LEX production process at the LHC (left) and at a muon collider (right)
\end{center}
However, the same type of process could happen at e-p or muon colliders. In particular, at a muon collider, the signal would be particularly interesting as it would have a jet (quark) in the final-state particles. This process, $l\overline{l}\rightarrow \nu q \phi$, is shown in the right panel of Fig. 4. It is also notable that, if the only interaction vertex is the one through which the LEX state was created, decay of the scalar would lead to an additional quark.

\section{Spin 1/2 LEX States}

In the case that the LEX field is a fermion and the operator contains no derivatives, the operator must contain a quark, the LEX state, and the electroweak field strength tensor $W_{\mu\nu}$. Operators with only these fields are dimension 5, so we can also include a Higgs field to increase the dimension of these operators to 6. 
For these operators involving the LEX fermion and no derivatives, the LEX state will always be a 3 under SU(3). This occurs because the only SM fields in these operators that have a non-singlet SU(3) charge will be single quarks. In the tables below, $\psi$ is used to indicate the LEX fermion state.

\subsection{Operators with $\boldsymbol{Q\psi W}$}
We list operators without a Higgs field present in Table 4. These dimension 5 operators contain a LEX state that is a triplet under SU(2) if there is a right-handed quark. Conversely, the LEX state follows the SU(2) product rule $3 \otimes 2 = 2 \oplus 4$ when there is a left-handed quark. In this case, the LEX state is a doublet or quadruplet under SU(2). 

\begin{center}
\begin{tabular}{|c|c|c|}
\hline 
dimension & Operators with fermion LEX fields & (SU(3),SU(2),Y)\tabularnewline
\hline 
\hline 
dim 5 & $\overline{Q_{L}}_{i}\sigma^{\mu \nu} \psi_j W_{\mu \nu}^{i j}$ & ($3,2,1/6$)\tabularnewline
\hline 
dim 5 & $\overline{Q_{L}}^{i}\sigma^{\mu \nu} \psi_{i j k} W_{\mu \nu}^{j k}$ & ($3,4,1/6$)\tabularnewline
\hline
dim 5 & $\overline{u}\sigma^{\mu \nu} \psi_{i j} W_{\mu \nu}^{i j}$ & ($3,3,2/3$)\tabularnewline
\hline 
dim 5 & $\overline{d}\sigma^{\mu \nu} \psi_{i j} W_{\mu \nu}^{i j}$ & ($3,3,-1/3$)\tabularnewline
\hline 
\end{tabular}
\\
Table 4. Operators containing $Q\psi W$
\end{center}

\subsection{Operators with $\boldsymbol{Q\psi WH}$}
Next, we examine the operators with both a fermionic LEX field and an SU(2) field strength tensor that also include a Higgs field. In this case, there are many options for the SU(2) charge of the LEX field, depending on whether it couples to a left- or right-handed SM quark. 
In the case of the right-handed quark field, the SU(2) multiplication works the same as for the operators containing a left-handed quark in Section 3.1: the LEX field is either a doublet or quadruplet under SU(2). For the left-handed quark, we follow the SU(2) multiplication rule $3 \otimes 2 \otimes 2 = 1 \oplus 3 \oplus 5$. The LEX state here can be a singlet, triplet, or quintuplet under SU(2).

\begin{center}
\begin{tabular}{|c|c|c|}
\hline 
dimension & Operators with fermion LEX fields & (SU(3),SU(2),Y)\tabularnewline
\hline 
\hline 
dim 6 & $\overline{Q_{L}}_{i}\sigma^{\mu \nu} \psi W_{\mu \nu}^{i j} H_j$ & ($3,1,-1/3$)\tabularnewline
\hline
dim 6 & $\overline{Q_{L}}_{i}\sigma^{\mu \nu} \psi W_{\mu \nu}^{i j} H^\dagger_j$ & ($3,1,2/3$)\tabularnewline
\hline
\multirow{2}{*}{dim 6} & $\overline{Q_{L}}_{i}\sigma^{\mu \nu} \psi^i_j W_{\mu \nu}^{j k} H_k$ & \multirow{2}{*}{($3,3,-1/3$)}\tabularnewline
& $\overline{Q_{L}}_{i}\sigma^{\mu \nu} \psi^k_j W_{\mu \nu}^{i j} H_k$ & \tabularnewline
\hline
\multirow{2}{*}{dim 6} & $\overline{Q_{L}}_{i}\sigma^{\mu \nu} \psi^i_j W_{\mu \nu}^{j k} H^\dagger_k$ & \multirow{2}{*}{($3,3,2/3$)}\tabularnewline
 & $\overline{Q_{L}}_{i}\sigma^{\mu \nu} \psi^k_j W_{\mu \nu}^{i j} H^\dagger_k$ & \tabularnewline
\hline
dim 6 & $\overline{Q_{L}}^{i}\sigma^{\mu \nu} \psi_{i j k l} W_{\mu \nu}^{j k} H^l$ & ($3,5,-1/3$)\tabularnewline
\hline 
dim 6 & $\overline{Q_{L}}^{i}\sigma^{\mu \nu} \psi_{i j k l} W_{\mu \nu}^{j k} H^{\dagger l}$ & ($3,5,2/3$)\tabularnewline
\hline 
dim 6 & $\overline{u}\sigma^{\mu \nu} \psi_i W_{\mu \nu}^{i j} H_j$ & ($3,2,1/6$)\tabularnewline
\hline
dim 6 & $\overline{u}\sigma^{\mu \nu} \psi_i W_{\mu \nu}^{i j} H^{\dagger}_j$ & ($3,2,7/6$)\tabularnewline
\hline
dim 6 & $\overline{u}\sigma^{\mu \nu} \psi_{i j k} W_{\mu \nu}^{i j} H^k$ & ($3,4,1/6$)\tabularnewline
\hline
dim 6 & $\overline{u}\sigma^{\mu \nu} \psi_{i j k} W_{\mu \nu}^{i j} H^{\dagger k}$ & ($3,4,7/6)$\tabularnewline
\hline

dim 6 & $\overline{d}\sigma^{\mu \nu} \psi_i W_{\mu \nu}^{i j} H_j$ & ($3,2,-5/6$)\tabularnewline
\hline
dim 6 & $\overline{d}\sigma^{\mu \nu} \psi_i W_{\mu \nu}^{i j} H^{\dagger}_j$ & ($3,2,1/6$)\tabularnewline
\hline
dim 6 & $\overline{d}\sigma^{\mu \nu} \psi_{i j k} W_{\mu \nu}^{i j} H^k$ & ($3,4,-5/6$)\tabularnewline
\hline
dim 6 & $\overline{d}\sigma^{\mu \nu} \psi_{i j k} W_{\mu \nu}^{i j} H^{\dagger k}$ & ($3,4,1/6$)\tabularnewline
\hline
\end{tabular}
\\
Table 5. Operators containing $Q\psi WH$
\end{center}

\subsection{Some Collider Phenomenology with Fermion LEX Field}

In comparison with the scalar LEX states, the fermionic LEX states outlined so far are all SU(3) fundamentals. However, they still access higher SU(2) representations, once again going all the way up to SU(2) quintuplets. 

Operators of these types can have 3, 4, or 5 particle interaction vertices. The three particle interactions can be found from operators in Table 4, or those in Table 5 when the Higgs is set to its vev. Four particle interactions can also be found in Table 4 or Table 5. The four particle interactions from Table 4 involve two gauge bosons, a quark, and the LEX fermion. The four particle interactions from Table 5 can contain the same mix of particles if the Higgs is set to its vev. Conversely, they can contain a single gauge boson, a Higgs boson, the quark, and the LEX fermion when the Higgs is not set to its vev. The five particle interactions contain two gauge bosons, one quark, a Higgs boson, and the LEX fermion. Each of these types of vertices are shown in Fig. 5.
\begin{center}
    \includegraphics[scale=.1]{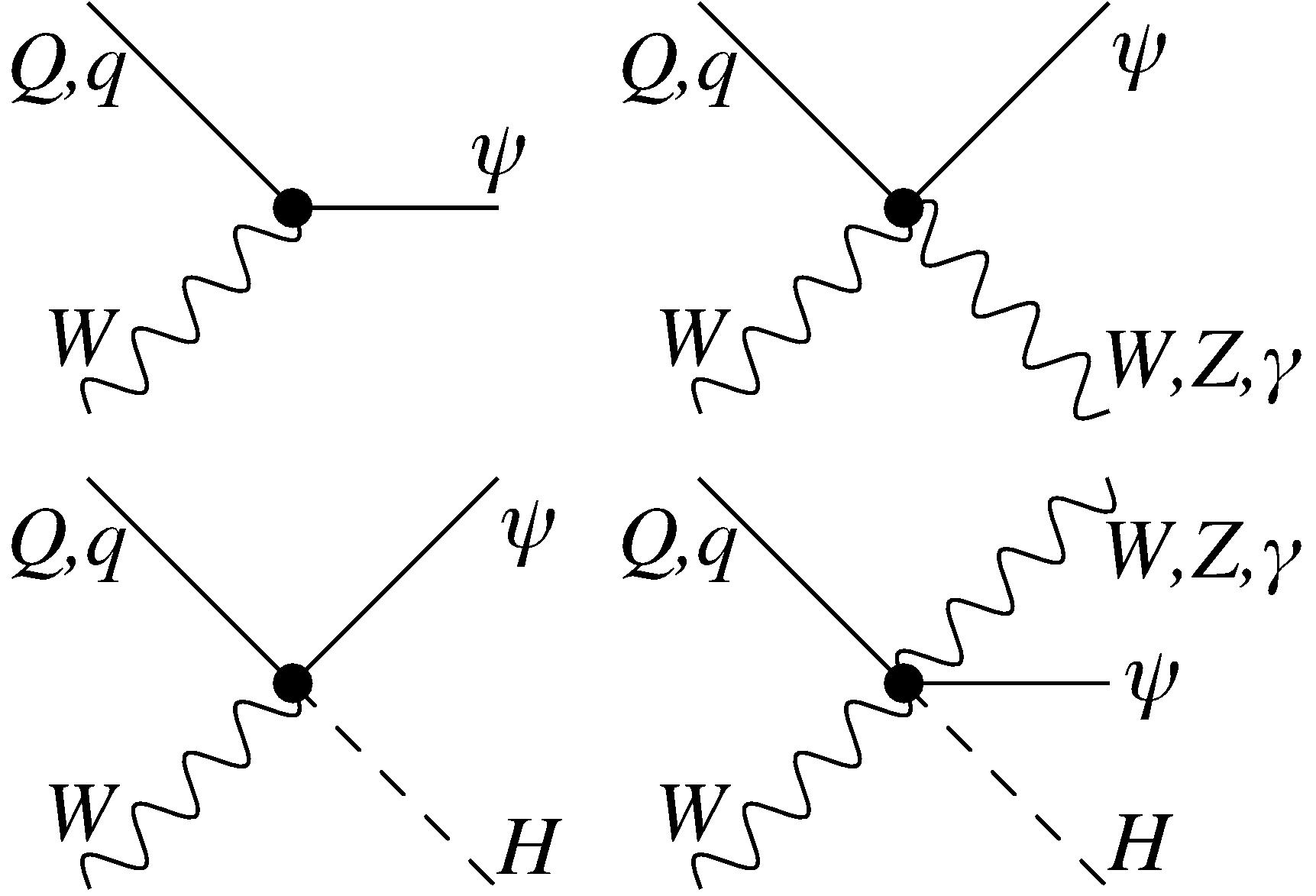} \\
    Figure 5. Interaction vertices involving fermionic LEX states, found in tables from Section 3
\end{center}

From these vertices, we can identify some interesting collider processes at the LHC. In particular, the process $q q \rightarrow \psi q$ shown in Fig. 6 would result in a very clean signature. Even if the fermion were to decay through the three particle interaction vertex, the resultant final state would contain only two jets, in addition to the decay product of a W boson (likely a lepton and MET). This is a very minimal set of final state particles.
\begin{center}
    \includegraphics[scale=.1]{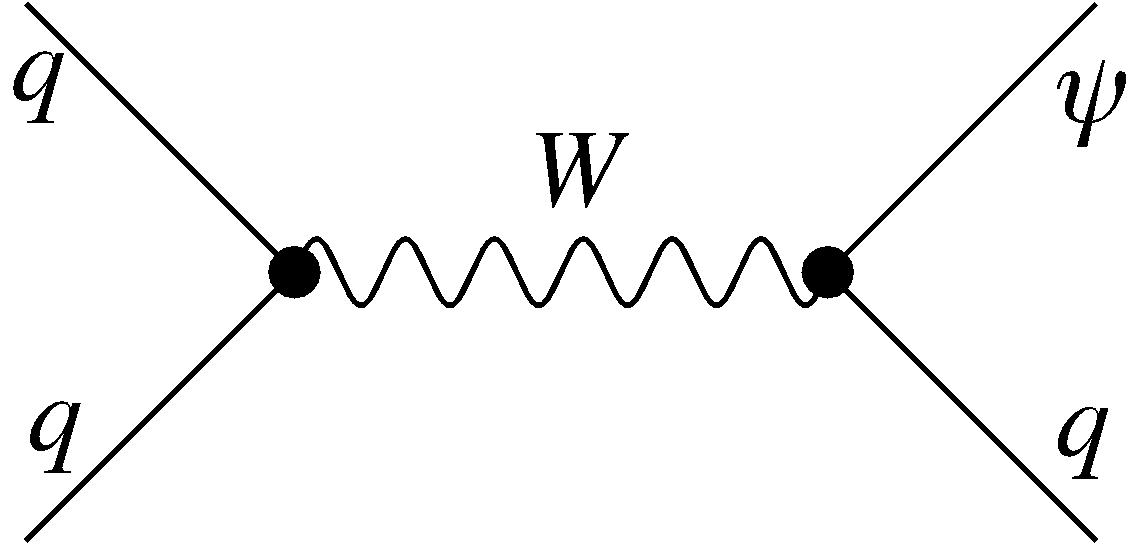}\\
    Figure 6. Potential $\psi$ production process at the LHC
\end{center}
Another process with the same final state particles ($qq\rightarrow\psi q$) involves a radiated W boson. This process, including the decay of the LEX state, is shown in Fig. 7. It would be interesting for the same reasons as the process shown in Fig. 6, but would likely have somewhat different kinematic distributions.

\begin{center}
    \includegraphics[scale=.1]{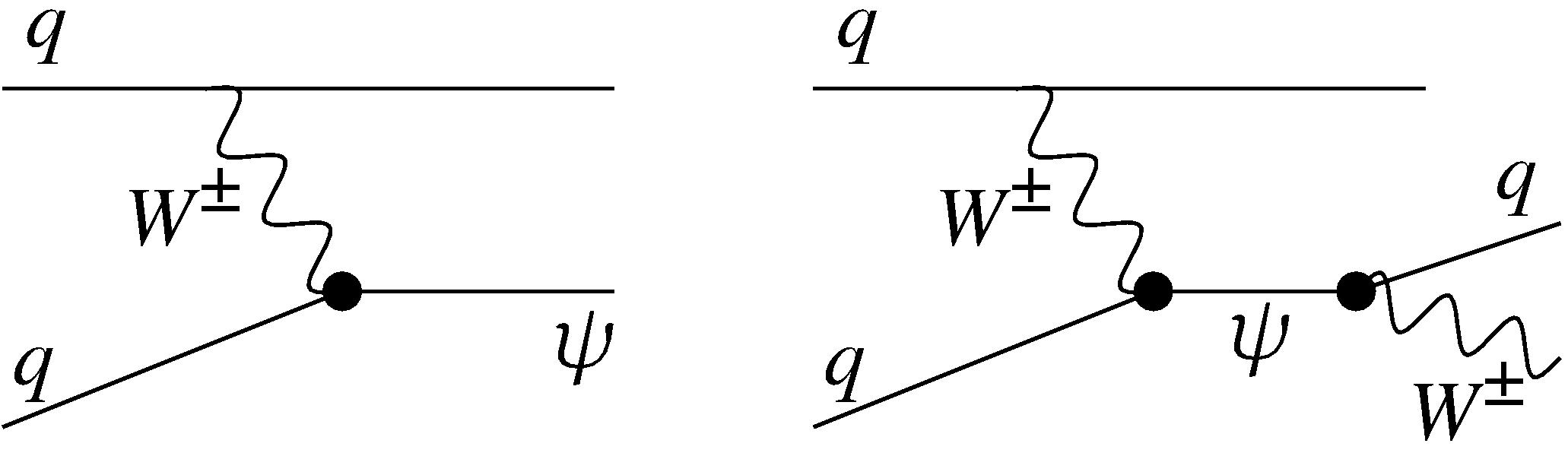}\\
    Figure 7. Potential $\psi$ production process at the LHC
\end{center}

\section{Derivatives with Scalar LEX Field}
In this section, we will look at operators which both contain a LEX scalar and include at least one covariant derivative. Operators of this type must include at least one quark, along with one other SM fermion, the LEX state, and one derivative. There may also be additional derivatives or Higgs fields involved in the operators. 
In this case, the W boson is generated through the covariant derivative acting on a non-singlet SU(2) state. Specifically, we know that 
\begin{equation}
    D^{\mu}\Phi \supset \partial^{\mu} \Phi + i g_3 \tau_3  A_3^{\mu}\Phi +i g_2 \tau_2 A_2^{\mu}\Phi  + ig_1 Y_{\phi} A_1^{\mu}\Phi \ ,
\end{equation}
where $\Phi$ is a field charged under each of the (SU(3), SU(2), SU(1)) groups. Here, the W boson is included in the $A_2$ gauge components. As such, in order for the covariant derivative to generate the W boson, we will require at least one of the SM fields in the following operators to carry non-singlet charge under SU(2).

\subsection{Operators with $\boldsymbol{DQf\phi}$}
We will start by examining operators with two SM fermions, at least one of which must be a quark, the LEX scalar, and one covariant derivative. In order to introduce non-singlet SU(2) charge, one of the SM fermions must be left-handed. 
In order for the operator to be Lorentz invariant, there must also be a gamma matrix present. With one derivative and three fields, we have 3 possible operators, which are related by one integration by parts relation.
\begin{equation}
    D_\mu \overline{f_1} \gamma^\mu f_2 \phi = \overline{f_1} \gamma^\mu D_\mu f_2 \phi + \overline{f_1} \gamma^\mu f_2 D_\mu \phi
\end{equation}
This brings us down to a maximum of two independent operators, in the case that all three fields present are charged under SU(2). In this case, we choose to drop the final operator, where the derivative is acting on the LEX field. However, in the case that one of the fields is an SU(2) singlet, an additional term in Eq. 3 is removed from the W-quark portal, as the derivative acting on that field would not generate the W boson. In this case, we only have one independent operator as part of the portal. The remaining operators of this type are listed in Table 6.

For the operators with only one quark, the LEX state will again be a 3 (or $\overline{3}$) under SU(3). For operators with a quark and an antiquark, the LEX state must be an SU(3) singlet or octet. In operators with a quark and the charge conjugate of an antiquark, the LEX state will be a triplet or sextet under SU(3). Furthermore, in operators with only one left-handed SM fermion, the LEX state will be an SU(2) doublet. Operators containing two left-handed SM fields follow the SU(2) product rule $2 \otimes 2 = 1 \oplus 3$, so the LEX state is an SU(2) singlet or triplet.

\begin{center}
\begin{tabular}{|c|c|c|}
\hline 
dimension & Operators with fermion LEX fields & (SU(3),SU(2),Y)\tabularnewline
\hline 
\hline 
dim 5 & $D_\mu \overline{Q_L}^i\gamma^\mu L_j \phi^j_i $, \quad $\overline{Q_L}^i\gamma^\mu D_\mu L_j  \phi^j_i $& ($3,3,2/3$)\tabularnewline
\hline
dim 5 & $D_\mu \overline{Q_L}^i\gamma^\mu L_i \phi $ & ($3,1,2/3$)\tabularnewline
\hline
dim 5 & $D_\mu \overline{Q_L^c}_i\gamma^\mu \ell \phi^i $ & ($\overline{3},2,5/6$)\tabularnewline
\hline
dim 5 & $\overline{u^c} \gamma^\mu D_\mu L_i  \phi^i$ & ($\overline{3},2,-1/6$) \tabularnewline
\hline
dim 5 & $\overline{d^c} \gamma^\mu D_\mu L_i  \phi^i$ & ($\overline{3},2,5/6$) \tabularnewline
\hline
dim 5 & $D_\mu \overline{Q_L}^i\gamma^\mu Q_j \phi^j_i $, \quad $ \overline{Q_L}^i\gamma^\mu D_\mu Q_j \phi^j_i $ & ($8,3,0$) , ($1,3,0$)\tabularnewline
\hline
dim 5 & $D_\mu \overline{Q_L}^i\gamma^\mu Q_i \phi $ & ($8,1,0$) , ($1,1,0$)\tabularnewline
\hline
dim 5 & $D_\mu \overline{Q_L^c}_i\gamma^\mu u \phi^i $ & ($\overline{6},2,-5/6$) , ($3,2,-5/6$) \tabularnewline
\hline
dim 5 & $D_\mu \overline{Q_L^c}_i\gamma^\mu d \phi^i $ & ($\overline{6},2,1/6$) , ($3,2,1/6$) \tabularnewline
\hline
\end{tabular}
\\
Table 6. Operators containing $DQf\phi$
\end{center}

It is notable that, if one plugs in fermion equations of motion, operators of this type would disappear from the portal. However, in including these operators, we are making a choice of basis. We are choosing to keep the maximal set of operators which can be included in the portal, with the understanding that the W boson is required to be off-shell.

\subsection{Operators with $\boldsymbol{DQf\phi H}$}
The next operator type in the W-quark portal that contains a scalar LEX field involves dimension six operators with one covariant derivative, the LEX field, a Higgs boson, and two SM fermions, at least one of which is a quark. The two fermions can be both left-handed, both right-handed, or a mix of left- and right-handed due to the SU(2) charge of the Higgs. Assuming that all of the fields are charged under SU(2), there are four possible operators, related by one integration by parts.
\begin{equation}
    D_\mu \overline{f_1} \gamma^\mu f_2 \phi H = \overline{f_1} \gamma^\mu D_\mu f_2 \phi H + \overline{f_1} \gamma^\mu f_2 D_\mu \phi H + \overline{f_1} \gamma^\mu f_2 \phi D_\mu H
\end{equation}
Here, we choose to drop the operator where the derivative acts on the LEX field. For operators with one or more SU(2) singlet fields, the number of operators further decreases. We list operators of this type in Tables 7 and 8.

\begin{center}
\begin{tabular}{|c|c|c|}
\hline 
dimension & Operators with fermion LEX fields & (SU(3),SU(2),Y)\tabularnewline
\hline 
\hline 
\multirow{2}{*}{dim 6} &  $D_\mu \overline{Q_L}^i \gamma^\mu Q_{Lj} \phi^{jk}_i H_k$, \quad $\overline{Q_L}^i \gamma^\mu D_\mu Q_{Lj} \phi^{jk}_i H_k$ & {($8,4,-1/2$) ,}\tabularnewline
  &  $\overline{Q_L}^i \gamma^\mu Q_{Lj} \phi^{jk}_i D_\mu H_k$ & ($1,4,-1/2$) \tabularnewline
\hline
\multirow{3}{*}{dim 6} &  $D_\mu \overline{Q_L}^i \gamma^\mu Q_{Li} \phi^{j} H_j$, \quad $D_\mu \overline{Q_L}^i \gamma^\mu Q_{Lj} \phi^{j} H_i$ & \multirow{2}{*}{($8,2,-1/2$) ,} \tabularnewline

 &  $\overline{Q_L}^i \gamma^\mu D_\mu Q_{Li} \phi^{j} H_j$, \quad $\overline{Q_L}^i \gamma^\mu D_\mu Q_{Lj} \phi^{j} H_i$ & \multirow{2}{*}{($1,2,-1/2$)}\tabularnewline

 &  $\overline{Q_L}^i \gamma^\mu Q_{Li} \phi^{j} D_\mu H_j$, \quad $\overline{Q_L}^i \gamma^\mu Q_{Lj}  \phi^{j} D_\mu H_i$ & \tabularnewline
\hline

\multirow{2}{*}{dim 6} &  $D_\mu \overline{Q_L}^i \gamma^\mu L_j \phi^{jk}_i H_k$, \quad $\overline{Q_L}^i \gamma^\mu D_\mu L_{j} \phi^{jk}_i H_k$ & \multirow{2}{*}{($3,4,1/6$)}\tabularnewline
 &  $\overline{Q_L}^i \gamma^\mu  L_{j} \phi^{jk}_i D_\mu H_k$ & \tabularnewline
\hline
\multirow{3}{*}{dim 6} &  $D_\mu \overline{Q_L}^i \gamma^\mu L_{i} \phi^{j} H_j$, \quad $D_\mu \overline{Q_L}^i \gamma^\mu L_{j} \phi^{j} H_i$ & \multirow{3}{*}{($3,2,1/6$)} \tabularnewline
 &  $\overline{Q_L}^i \gamma^\mu D_\mu L_{i} \phi^{j} H_j$, \quad $\overline{Q_L}^i \gamma^\mu D_\mu L_j \phi^{j} H_i$ &  \tabularnewline
 &  $\overline{Q_L}^i \gamma^\mu {L_i}  \phi^{j} D_\mu H_j$, \quad $\overline{Q_L}^i \gamma^\mu L_j \phi^{j} D_\mu H_i$ & \tabularnewline
\hline

\multirow{2}{*}{dim 6} &  $D_\mu \overline{Q_L}^i \gamma^\mu Q_{Lj} \phi^{j}_{ik} H^{\dagger k}$, \quad $\overline{Q_L}^i \gamma^\mu D_\mu Q_{Lj} \phi^{j}_{ik} H^{\dagger k}$  & \multirow{2}{*}{($8,4,1/2$) , ($1,4,1/2$)}\tabularnewline
 &  $\overline{Q_L}^i \gamma^\mu Q_{Lj} \phi^{j}_{ik} D_\mu H^{\dagger k}$ & \tabularnewline
\hline

\multirow{3}{*}{dim 6} &  $D_\mu \overline{Q_L}^i \gamma^\mu Q_{Li} \phi_{j} H^{\dagger j}$, \quad $D_\mu \overline{Q_L}^i \gamma^\mu Q_{Lj} \phi^{j} H^\dagger_i$ & \multirow{3}{*}{($8,2,1/2$) , ($1,2,1/2$)}\tabularnewline
&  $\overline{Q_L}^i \gamma^\mu D_\mu Q_{Li} \phi_{j} H^{\dagger j}$, \quad $\overline{Q_L}^i \gamma^\mu D_\mu Q_{Lj} \phi^{j} H^\dagger_i$ & \tabularnewline
 &  $\overline{Q_L}^i \gamma^\mu Q_{Li} \phi_{j} D_\mu H^{\dagger j}$, \quad $\overline{Q_L}^i \gamma^\mu Q_{Lj} \phi^{j} D_\mu H^\dagger_i$ & \tabularnewline
\hline

\multirow{2}{*}{dim 6} &  $D_\mu \overline{Q_L}^i \gamma^\mu L_j \phi^{j}_{ik} H^{\dagger k}$, \quad $\overline{Q_L}^i \gamma^\mu D_\mu L_{j} \phi^{j}_{ik} H^{\dagger k}$ & \multirow{2}{*}{($3,4,7/6$)}\tabularnewline
 &  $\overline{Q_L}^i \gamma^\mu L_{j} \phi^{j}_{ik} D_\mu H^{\dagger k}$ & \tabularnewline
\hline
\multirow{3}{*}{dim 6} &  $D_\mu \overline{Q_L}^i \gamma^\mu L_{i} \phi_{j} H^{\dagger j}$, \quad $D_\mu \overline{Q_L}^i \gamma^\mu L_{j} \phi^{j} H^\dagger_i$ & \multirow{3}{*}{($3,2,7/6$)} \tabularnewline
&  $\overline{Q_L}^i \gamma^\mu D_\mu L_{i} \phi_{j} H^{\dagger j}$, \quad $\overline{Q_L}^i \gamma^\mu D_\mu L_j \phi^{j} H^\dagger_i$ & \tabularnewline
 &  $\overline{Q_L}^i \gamma^\mu {L_i} \phi_{j} D_\mu H^{\dagger j}$, \quad $\overline{Q_L}^i \gamma^\mu L_j \phi^{j} D_\mu H^\dagger_i$ & \tabularnewline
\hline

dim 6 &  $D_\mu \overline{Q_L^c}_i \gamma^\mu  \ell \phi^{ij} H_j$, \quad $\overline{Q_L^c}_i \gamma^\mu \ell \phi^{ij} D_\mu H_j$ & ($\overline{3},3,1/3$)\tabularnewline
\hline
dim 6 &  $D_\mu \overline{Q_L^c}_i \gamma^\mu  \ell \phi H^i$ &($\overline{3},1,1/3$)\tabularnewline
\hline

dim 6 &  $D_\mu \overline{Q_L^c}_i \gamma^\mu  \ell \phi^{i}_j H^{\dagger j}$, \quad $\overline{Q_L^c}_i \gamma^\mu \ell \phi^{i}_j D_\mu H^{\dagger j}$ & ($\overline{3},3,4/3$)\tabularnewline
\hline
dim 6 &  $D_\mu \overline{Q_L^c}_i \gamma^\mu  \ell \phi H^{\dagger i}$, \quad $\overline{Q_L^c}_i \gamma^\mu \ell \phi D_\mu H^{\dagger i}$ & ($\overline{3},1,4/3$) \tabularnewline
\hline
dim 6 &  $\overline{u^c} \gamma^\mu D_\mu Q_{Li} \phi^{ij} H_j$, \ $\overline{u^c} \gamma^\mu Q_{Li} \phi^{ij} D_\mu H_j$ & ($\overline{6},3,-4/3$), ($3,3,-4/3$)\tabularnewline
\hline
dim 6 &  $\overline{d^c} \gamma^\mu D_\mu Q_{Li} \phi^{ij} H_j$, \ $\overline{d^c} \gamma^\mu Q_{Li} \phi^{ij} D_\mu H_j$ & ($\overline{6},3,-1/3$), ($3,3,-1/3$)\tabularnewline
\hline

dim 6 &  $\overline{u^c} \gamma^\mu D_\mu Q_{Li} \phi^{} H^i$ & ($\overline{6},1,-4/3$), ($3,1,-4/3$)\tabularnewline
\hline
dim 6 &  $\overline{d^c} \gamma^\mu D_\mu Q_{Li} \phi^{} H^i$ & ($\overline{6},1,-1/3$), ($3,1,-1/3$)\tabularnewline
\hline

dim 6 &  $\overline{u^c} \gamma^\mu D_\mu Q_{Li} \phi^{i}_j H^{\dagger j}$, \quad $\overline{u^c} \gamma^\mu Q_{Li} \phi^{i}_j D_\mu H^{\dagger j}$ & ($\overline{6},3,-1/3$), ($3,3,-1/3$)\tabularnewline
\hline
dim 6 &  $\overline{d^c} \gamma^\mu D_\mu Q_{Li} \phi^{i}_j H^{\dagger j}$, \quad $\overline{d^c} \gamma^\mu Q_{Li} \phi^{i}_j D_\mu H^{\dagger j}$ & ($\overline{6},3,2/3$), ($3,3,2/3$)\tabularnewline
\hline

dim 6 &  $\overline{u^c} \gamma^\mu D_\mu Q_{Li} \phi^{} H^{\dagger i}$ & ($\overline{6},1,-1/3$), ($3,1,-1/3$)\tabularnewline
\hline
dim 6 &  $\overline{d^c} \gamma^\mu D_\mu Q_{Li} \phi^{} H^{\dagger i}$ & ($\overline{6},1,2/3$), ($3,1,2/3$)\tabularnewline
\hline

\end{tabular}
\\
Table 7. Operators containing $DQf \phi H$
\end{center}

\begin{center}
\begin{tabular}{|c|c|c|}
\hline 
dimension & Operators with fermion LEX fields & (SU(3),SU(2),Y)\tabularnewline
\hline 
\hline

dim 6 &  $\overline{u^c} \gamma^\mu D_\mu L_{i} \phi^{ij} H_j$, \quad $\overline{u^c} \gamma^\mu L_{i} \phi^{ij} D_\mu H_j$ & ($\overline{3},3,-2/3$)\tabularnewline
\hline
dim 6 &  $\overline{d^c} \gamma^\mu D_\mu L_{i} \phi^{ij} H_j$, \quad $\overline{d^c} \gamma^\mu L_{i} \phi^{ij} D_\mu H_j$ & ($\overline{3},3,1/3$)\tabularnewline
\hline

dim 6 &  $\overline{u^c} \gamma^\mu D_\mu L_{i} \phi^{} H^i$& ($\overline{3},1,-2/3$) \tabularnewline
\hline
dim 6 &  $\overline{d^c} \gamma^\mu D_\mu L_{i} \phi^{} H^i$& ($\overline{3},1,1/3$) \tabularnewline
\hline

dim 6 &  $\overline{u^c} \gamma^\mu D_\mu L_{i} \phi^{i}_j H^{\dagger j}$, \quad $\overline{u^c} \gamma^\mu L_{i} \phi^{i}_j D_\mu H^{\dagger j}$ & ($\overline{3},3,1/3$)\tabularnewline
\hline
dim 6 &  $\overline{d^c} \gamma^\mu D_\mu L_{i} \phi^{i}_j H^{\dagger j}$, \quad $\overline{d^c} \gamma^\mu L_{i} \phi^{i}_j D_\mu H^{\dagger j}$ & ($\overline{3},3,4/3$)\tabularnewline
\hline

dim 6 &  $\overline{u^c} \gamma^\mu D_\mu L_{i} \phi^{} H^{\dagger i}$& ($\overline{3},1,1/3$) \tabularnewline
\hline
dim 6 &  $\overline{d^c} \gamma^\mu D_\mu L_{i} \phi^{} H^{\dagger i}$ & ($\overline{3},1,4/3$) \tabularnewline
\hline

dim 6 &  $\overline{u} \gamma^\mu \ell \phi^i D_\mu H_i$& ($3,2,7/6$) \tabularnewline
\hline
dim 6 &  $\overline{d} \gamma^\mu \ell \phi^i D_\mu H_i$& ($3,2,1/6$) \tabularnewline
\hline
dim 6 &  $\overline{u} \gamma^\mu u \phi^i D_\mu H_i$& ($8,2,-1/2$), ($1,2,-1/2$) \tabularnewline
\hline
dim 6 &  $\overline{u} \gamma^\mu d \phi^i D_\mu H_i$& ($8,2,1/2$) , ($1,2,1/2$) \tabularnewline
\hline
dim 6 &  $\overline{d} \gamma^\mu d \phi^i D_\mu H_i$& ($8,2,-1/2$), ($1,2,-1/2$) \tabularnewline
\hline

dim 6 &  $\overline{u} \gamma^\mu \ell \phi_i D_\mu H^{\dagger i}$& ($3,2,13/6$) \tabularnewline
\hline
dim 6 &  $\overline{d} \gamma^\mu \ell \phi_i D_\mu H^{\dagger i}$& ($3,2,7/6$) \tabularnewline
\hline
dim 6 &  $\overline{u} \gamma^\mu u \phi_i D_\mu H^{\dagger i}$& ($8,2,1/2$) , ($1,2,1/2$) \tabularnewline
\hline
dim 6 &  $\overline{u} \gamma^\mu d \phi_i D_\mu H^{\dagger i}$& ($8,2,3/2$) , ($1,2,3/2$) \tabularnewline
\hline
dim 6 &  $\overline{d} \gamma^\mu d \phi_i D_\mu H^{\dagger i}$& ($8,2,1/2$) , ($1,2,1/2$) \tabularnewline
\hline

\end{tabular}
\\
Table 8. Operators containing $Dqf \phi H$
\end{center}

\subsubsection{Operators with $\boldsymbol{DDQf\phi}$}
Next, we consider operators with two SM fermions, the LEX scalar, and two derivatives. In order for a derivative to generate a W boson, at least one of the SM fermions must be left-handed. For these operators, we must consider all possible forms that the two derivatives can take. As the Lorentz indices must be contracted, the following three forms are available.
\begin{equation}
    D^\mu D_\mu , \quad D^{\mu} \sigma_{\mu \nu}D^\nu , \quad\slashed{D}\slashed{D}    
\end{equation}
We will start by addressing how any two of these derivative structures can be eliminated. This can be done by proving  both their redundancy with other, previously discussed operators and their redundancy with one another.

We begin with the $D^{\mu} \sigma_{\mu \nu}D^\nu$ structure. This structure has six possible operators, related by three integrations by parts relations:
\begin{align}
\begin{split}
    D_{\mu}D_\nu \overline{f_1}\sigma^{\mu\nu} f_2 \phi = D_\nu\overline{f_1} \sigma^{\mu\nu}D_\mu f_2 \phi + D_\nu\overline{f_1} \sigma^{\mu\nu}f_2 D_\mu \phi 
    \\
    \overline{f_1} \sigma^{\mu\nu} D_\mu D_\nu f_2 \phi = D_\mu\overline{f_1} \sigma^{\mu\nu}D_\nu f_2 \phi + \overline{f_1}\sigma^{\mu\nu} D_\nu f_2 D_\mu \phi     
    \\
    \overline{f_1} \sigma^{\mu\nu}f_2 D_\mu D_\nu \phi = D_\mu\overline{f_1}\sigma^{\mu\nu} f_2 D_\nu \phi + \overline{f_1}\sigma^{\mu\nu} D_\mu f_2 D_\nu \phi     
\end{split}
\end{align}
This results in three independent operators. We can choose the remaining operators to be the ones where both derivatives act on the same field: $D_\mu D_\nu \overline{f_1}\sigma^{\mu \nu} f_2 \phi$, $\overline{f_1}\sigma^{\mu \nu} D_\mu D_\nu f_2 \phi$, and $\overline{f_1}\sigma^{\mu \nu} f_2 D_\mu D_\nu \phi$. 

We also know the relation
\begin{equation}
    D_\mu D_\nu = \frac{g}{2} F_{\mu \nu} + \frac{1}{2} \{D_\mu , D_\nu\} \ .
\end{equation}
Contracting the antisymmetric $\sigma^{\mu \nu}$ with the anticommutator of the covariant derivatives results in zero, leaving us with only the first term in the right hand side of Eq. 11. Plugging this into the remaining operators, we are left with $W_{\mu \nu} \overline{f_1}\sigma^{\mu \nu} f_2 \phi$ as the contribution to the W-quark portal, which has the same form as the operators in Section 2. As such, this form is redundant.

Next, we can examine the operators that include derivative forms with two covariant derivatives and two gamma matrices ($\slashed{D} \slashed{D}$). 
Again, we have three independent operator choices, which we choose as the two derivatives acting on the same field. However, we know that for fermions $f$,
\begin{equation}
    \slashed{D} \slashed{D}f = D^2f + i\sigma^{\mu\nu} F_{\mu\nu} f \ .
\end{equation}
This relation also holds for scalar particles, with the caveat that the gamma matrices contract with fermion spin indices.
We note that the operator form of the final term in Eq. 12 has already been addressed in the portal. 

The final Lorentz structure that we need to discuss for the double-derivative structure is $D\mu D^\mu=D^2$. This can be seen as the remaining non-redundant term on the right-hand side of Eq. 7. As such, there is only one independent Lorentz structure with two derivatives. In the following discussion, we choose to work with the two-derivative Lorentz structure $\slashed{D}\slashed{D}$.

As previously mentioned, there are six possible operators with two SM fermions, a LEX scalar, and the $\slashed{D}\slashed{D}$ derivative structure. Due to equations of motion, this was decreased to three independent operators. However, the specific three operators that we chose are not the mandatory choice. While in the previous discussion it was illuminating to consider operators where both derivatives acted on the same field, other choices are equally valid. For the following discussion, we shift our choice to a slightly different set of three operators: 
\begin{equation}
    \slashed{D}\slashed{D}\overline{f_1}f_2\phi, \quad \slashed{D}\overline{f_1}\slashed{D}f_2\phi, \quad 
    \overline{f_1}\slashed{D}\slashed{D}f_2\phi \ .
\end{equation}

With the derivatives acting only on SM fermion fields, we can use the equations of motion to get rid of one of the derivatives in each of the operators listed in Eq. 13. This essentially allows us to switch out left- and right-handed SM fermions, and to gain a Higgs field in place of the derivative. These equations of motion are
\begin{align}
    \begin{split}
        \slashed{D}\overline{Q_L} &= y_d \overline{d_R} H + y_u \overline{u_R} H^\dagger
        \\
        \slashed{D}\overline{u_R} &= y_u\overline{Q_L}H
        \\
        \slashed{D}\overline{d_R} &= y_d\overline{Q_L}H^\dagger
        \\
        \slashed{D}L &= y_\ell \ell H
        \\
        \slashed{D}\ell &= y_\ell L H^\dagger \ .
    \end{split}
\end{align}
As such, the remaining operators of this type transform to those of type $DQfH\phi$, which has already been added to the portal in Section 4.2. There are no new operators of this type.

\subsection{Some Collider Phenomenology with Scalar LEX Fields and Derivatives}
We find a similar range of SU(3) quantum numbers as in Section 2. However, the SU(2) values are slightly more restrictive, only reaching quadruplets here. Still, this leaves us with a good number of unique quantum numbers and charge availability. Looking at an example field, we choose one with a quadruplet SU(2) value and octet SU(3) charge. This field, found in the dimension 6 operator $\frac{1}{\Lambda^2}D_\mu \overline{Q_L}^i \gamma^\mu Q_{Lj} \phi^{jk}_i H_k$, has charged components $(1,0,-1,-2)$.

The same interaction vertices as those shown in Fig. 2 also apply to operators in this section of the paper. As such, the interactions shown in Fig. 4 could also be a result of the operators with derivatives. However, the operators with derivatives add 5 particle interactions which contain a quark, W boson, Higgs, an additional SM fermion, and the LEX scalar. This leads to collider processes such as the one shown in Fig. 8. 
\begin{center}
    \includegraphics[scale=.1]{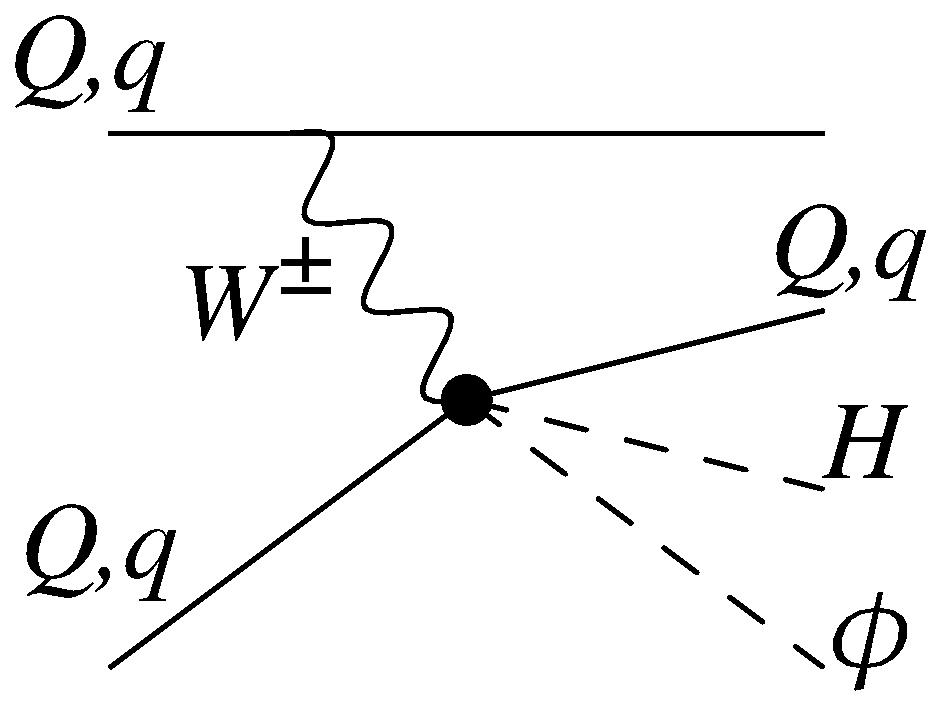} \\
    Figure 8. Associated $\phi$ production resulting from interactions listed in Section 4.2
\end{center}
However, when there are more final-state particles, the accessible LEX states cannot have as high of a mass. Additionally, signals such as the one shown in Fig. 8 can be much messier. As such, processes like this one are not the most likely to discover new states.

\section{Derivatives with fermionic LEX Field}
Next, we will examine operators which contain a LEX fermion and include at least one covariant derivative. Operators of this type must include the quark, the LEX state, and one derivative. There may also be additional derivatives, Higgs fields, or a field strength tensor involved in the operators.

\subsection{Containing derivatives and a field strength tensor}

\subsubsection{Operators with $\boldsymbol{Q D\psi W}$ and $\boldsymbol{q D\psi W}$}
We start by discussing operators that include a quark, LEX fermion, W field strength tensor, and derivative. In this case, the quark may be left- or right-handed. In either form, the derivative can act on any of the fields. Starting with the case of a left-handed quark, we have the operators below.
\begin{equation}
    \overline{Q}\gamma^\mu \overleftarrow{D^\nu} \psi W_{\mu \nu},
    \quad \overline{Q}\gamma^\mu \overrightarrow{D^\nu} \psi W_{\mu \nu},
    \quad \overline{Q}\gamma^\mu \psi D^\nu W_{\mu \nu}
\end{equation}
However, these are related by integration by parts:
\begin{equation}
    \overline{Q}\gamma^\mu \overleftarrow{D^\nu} \psi W_{\mu \nu} = 
    \overline{Q}\gamma^\mu \overrightarrow{D^\nu} \psi W_{\mu \nu} +
    \overline{Q}\gamma^\mu \psi D^\nu W_{\mu \nu} \ .
\end{equation}

We are thus left with only two independent operators of this type, shown in Table 10. Each of the operators here contains one quark, making the LEX state a 3 under SU(3). For the operators containing a left-handed quark, we follow the SU(2) multiplication rule $2 \otimes 3 = 2 \oplus 4$, so the LEX state will be a doublet or quadruplet under SU(2). For the operators with a right-handed quark, the LEX state must balance out the W field strength tensor alone, making it an SU(2) triplet.

\begin{center}
\begin{tabular}{|c|c|c|}
\hline 
dimension & Operators with fermion LEX fields & (SU(3),SU(2),Y)\tabularnewline
\hline 
\hline 
dim 6 & $\overline{Q}_{L i}\gamma^\mu \overleftarrow{D^\nu} \psi_j W_{\mu \nu}^{i j}$, \quad $\overline{Q}_{L i}\gamma^\mu \overrightarrow{D^\nu} \psi_j W_{\mu \nu}^{i j}$ & ($3,2,1/6$)\tabularnewline
\hline 
dim 6 & $\overline{Q}_{L }^{i}\gamma^\mu \overleftarrow{D^\nu} \psi_{i j k} W_{\mu \nu}^{j k}$, \quad $\overline{Q}_{L}^{ i}\gamma^\mu \overrightarrow{D^\nu} \psi_{i j k} W_{\mu \nu}^{jk}$ & ($3,4,1/6$)\tabularnewline
\hline 
dim 6 & $\overline{u}\gamma^\mu \overrightarrow{D^\nu} \psi_{ij} W_{\mu \nu}^{i j}$ & ($3,3,2/3$)\tabularnewline
\hline 
dim 6 & $\overline {d}\gamma^\mu \overrightarrow{D^\nu} \psi_{ij} W_{\mu \nu}^{i j}$ & ($3,3,-1/3$)\tabularnewline 
\hline
\end{tabular}
\\
Table 9. Operators containing $Q D\psi W$ and $q D\psi W$
\end{center}

\subsubsection{Operators with $\boldsymbol{QD\psi G}$ or $\boldsymbol{QD\psi B}$}
For operators containing a quark, LEX fermion, covariant derivative, and field strength tensor, the coupling to the W boson is produced by the derivative acting on a state that is charged under SU(2). As such, the quark in these operators must be left-handed. This would result in the LEX state being an SU(2) doublet. The derivative in these operators must act on the quark or LEX fermion. 
Through integration by parts relations, we know that 
\begin{equation}
    D^\nu \overline{Q}\gamma^\mu \psi F_{\mu \nu} = \overline{Q}\gamma^\mu D^\nu \psi F_{\mu \nu} + \overline{Q}\gamma^\mu \psi D^\nu F_{\mu \nu} \quad .
\end{equation}

However, as we know the last term does not contribute to the W-quark portal, we obtain $D^\nu \overline{Q}\gamma^\mu \psi F_{\mu \nu} = \overline{Q}\gamma^\mu D^\nu \psi F_{\mu \nu}$, leaving us with one remaining independent operator of this type. Due to the SU(3) field strength tensors, we are able to access operators of this type with 

\begin{center}
\begin{tabular}{|c|c|c|}
\hline 
dimension & Operators with fermion LEX fields & (SU(3),SU(2),Y)\tabularnewline
\hline 
\hline 
\multirow{2}{*}{dim 6} & \multirow{2}{*}{$\overline{Q_L}^i\gamma^\mu \overleftarrow{D^\nu} \psi_i G_{\mu \nu}$} & ($15,2,1/6$) , ($\overline{6},2,1/6$) ,\tabularnewline
 & & ($3,2,1/6$) \tabularnewline
\hline 
dim 6 & $\overline{Q_L}^i\gamma^\mu \overleftarrow{D^\nu} \psi_i B_{\mu \nu}$ & ($3,2,1/6$)\tabularnewline
\hline 
\end{tabular}
\\
Table 10. Operators containing $Q D\psi G$ and $Q D\psi B$
\end{center}

\subsection{Operators with $\boldsymbol{{Q}  D\psi}$}
We will now consider operators with a fermionic LEX state, at least one derivative, and no field strength tensors. We start with operators that contain a quark, LEX state and only one derivative. The quark must be left-handed to allow SU(2) non-singlet charge. We have two options for the operator, as the derivative can act on the quark or the LEX fermion. Integration by parts relates these two operators directly, leaving us with one unique operator: $\slashed{D} \overline{Q_L} \psi$.

\begin{center}
\begin{tabular}{|c|c|c|}
\hline 
dimension & Operators with fermion LEX fields & (SU(3),SU(2),Y)\tabularnewline
\hline 
\hline 
dim 4 & $\slashed{D} \overline{Q_L}^i \psi_i$ & ($3,2,1/6$)\tabularnewline
\hline 
\end{tabular}
\\
Table 11. Operators containing $Q D\psi$
\end{center}

\subsection{Operators with $\boldsymbol{DQ\psi H}$}
Next, we will look at dimension 5 operators containing one derivative and one Higgs boson. Due to the SU(2) charge of the Higgs, the quark can be either left- or right-handed. Starting with the case of a left-handed quark, the LEX particle is either a singlet or triplet under SU(2). In the case of the SU(2) triplet, the derivative could act on any of the fields to generate the W particle: 
\begin{equation}
    \overline{Q}\gamma^\mu \overleftarrow{D_\mu} \psi H \ , \quad  \overline{Q}\gamma^\mu \overrightarrow{D_\mu} \psi H \ , \quad  \overline{Q}\gamma^\mu  \psi {D_\mu} H \ .
\end{equation}
There is one integration by parts relation, leaving us with two terms.

For the case of the LEX SU(2) singlet, the derivative acting on the LEX state will not contribute to the portal. Similarly, for the case of a right-handed quark, the term where the derivative acts on the quark does not contribute to the portal. In this case, the LEX state will be an SU(2) doublet. In each of these cases, integration by parts leaves us with only one remaining operator.

\begin{center}
\begin{tabular}{|c|c|c|}
\hline 
dimension & Operators with fermion LEX fields & (SU(3),SU(2),Y)\tabularnewline
\hline 
\hline 
dim 5 & $\overline{Q_L}^i\gamma^\mu \overleftarrow{D_\mu} \psi^j_i H_j$, \quad $\overline{Q_L}^i \gamma^\mu \psi^j_i D_\mu H_j$ & ($3,3,-1/3$)\tabularnewline
\hline 
dim 5 & $\overline{Q_L}^i\gamma^\mu \overleftarrow{D_\mu} \psi H_i$ & ($3,1,-1/3$)\tabularnewline
\hline 
dim 5 & $\overline{u}\gamma^\mu  \psi^i D_\mu H_i$ & ($3,2,1/6$)\tabularnewline
\hline 
dim 5 & $\overline{d}\gamma^\mu  \psi^i D_\mu H_i$ & ($3,2,-5/6$)\tabularnewline
\hline 
dim 5 & $\overline{Q_L}^i\gamma^\mu \overleftarrow{D_\mu} \psi_{ij} H^{\dagger j}$, \quad $\overline{Q_L}^i \gamma^\mu  \psi_{ij} D_\mu H^{\dagger j}$ & ($3,3,2/3$)\tabularnewline
\hline 
dim 5 & $\overline{Q_L}^i\gamma^\mu \overleftarrow{D_\mu} \psi H^{\dagger}_i$ & ($3,1,2/3$)\tabularnewline
\hline 
dim 5 & $\overline{u}\gamma^\mu \psi_i D_\mu H^{\dagger i}$ & ($3,2,7/6$)\tabularnewline
\hline 
dim 5 & $\overline{d}\gamma^\mu \psi_i D_\mu H^{\dagger i}$ & ($3,2,1/6$)\tabularnewline
\hline 
\end{tabular}
\\
Table 12. Operators containing $Q D\psi H$
\end{center}

\subsubsection{Operators with $\boldsymbol{{Q} D D\psi}$}
In the following discussion, we examine operators with one quark, two covariant derivatives, and the LEX fermionic state. At first glance, we know that the quark must be left-handed to introduce non-singlet SU(2) charge. We also know that there are three possible Lorentz structures for the double derivatives.
However, the same reasoning from Section 4.2.1 applies here. Because the operator type including $Q\psi W_{\mu\nu}\sigma^{\mu\nu}$ has already been added to the portal in Section 3.1, we can reduce the three operator types to only one new, non-redundant option. Here, we choose to discuss the operators with the structure $\slashed{D}\slashed{D}$.

There are three operators with the $\slashed{D}\slashed{D}$ format:

\begin{equation}
    \slashed{D}\slashed{D} \overline{Q_L} \psi , \quad \overline{Q_L} \slashed{D}\slashed{D} \psi , 
    \quad \slashed{D} \overline{Q}\slashed{D} \psi
\end{equation}

These are related by two integration by parts equations, leaving only one operator with $\slashed{D}\slashed{D}$. Here, we choose to work with $\slashed{D}\slashed{D} \overline{Q_L}^i \psi_i$. However, after applying the fermion equations of motion, this operator is also redundant:
\begin{align}
\begin{split}
    \slashed{D}\slashed{D} \overline{Q_L}^i \psi_i &= \slashed{D}( y_d \overline{d_R} H + y_u \overline{u_R} H^\dagger)\psi_i \\
    &= y_d \slashed{D}\overline{d_R} H^i \psi_i + y_d \overline{d_R} \slashed{D} H^i  \psi_i+ y_u \slashed{D}\overline{u_R} H^\dagger\psi_i +  y_u \overline{u_R} \slashed{D}H^\dagger\psi_i
\end{split}
\end{align}
The first and third terms on the second line of Eq. 20 do not contribute to the W-quark portal, while the second and fourth terms are already included in Table 12. There are no new operators of this type.

\subsection{Operators with $\boldsymbol{DQ\psi HH}$}
Next, we examine operators containing one quark, the LEX fermion state, two Higgs bosons, and one covariant derivative. Operators of this type could contain left- or right-handed quarks, as the Higgs field is an SU(2) doublet and if the derivative acts on it we will generate a W boson. We will start by looking at the maximal case where all of the states within the operator have non-singlet SU(2) charge. This means that the quark is right-handed, and the LEX state has SU(2) numbers of $2\otimes2\otimes2 = (1\oplus3)\otimes2 = 2\oplus4$. In this case, there are 4 possible operators, and one integration by parts relation.
\begin{equation}
    D_\mu Q \gamma^\mu \psi HH= Q \gamma^\mu D_\mu\psi HH + Q \gamma^\mu \psi D_\mu HH + Q \gamma^\mu \psi H D_\mu H
\end{equation}
This leaves us with 3 operators. We choose to keep
\begin{equation}
    D_\mu Q \gamma^\mu \psi HH, \quad Q \gamma^\mu D_\mu \psi  H H, \ \quad Q \gamma^\mu \psi D_\mu H H \ .
\end{equation}
In the cases where there are two identical Higgs fields, this number further decreases to two.

In the case that the quark is right-handed, the LEX state is either an SU(2) triplet or singlet. Assuming the LEX state is a triplet and the two Higgs are identical, the LEX state must be a triplet and we will be left with only one operator. All operators with one quark, the LEX fermion, two Higgs bosons, and one derivative are shown in Table 13.

\begin{center}
\begin{tabular}{|c|c|c|}
\hline 
dimension & Operators with fermion LEX fields & (SU(3),SU(2),Y)\tabularnewline
\hline 
\hline 
dim 6 & $D_\mu \overline{Q_L}^i \gamma^\mu \psi_i^{jk} H_jH_k$ \ , \ $\overline{Q_L}^i \gamma^\mu \psi_i^{jk} H_jD_\mu H_k$ & ($3,4,-5/6$)\tabularnewline
\hline
dim 6 & $D_\mu \overline{Q_L}^i \gamma^\mu \psi^j H_iH_j$ \ , \ $\overline{Q_L}^i \gamma^\mu \psi^j H_iD_\mu H_j$ & ($3,2,-5/6$)\tabularnewline
\hline

\multirow{2}{*}{dim 6} & $D_\mu \overline{Q_L}^i \gamma^\mu \psi_{ij}^{k} H^{\dagger j}H_k$ \ , \ $ \overline{Q_L}^i \gamma^\mu D_\mu \psi_{ij}^{k} H^{\dagger j}H_k$ & \multirow{2}{*}{($3,4,1/6$)}\tabularnewline
 & $\overline{Q_L}^i \gamma^\mu \psi_{ij}^{k} H^{\dagger j}D_\mu H_k$ & \tabularnewline
\hline
    \multirow{4}{*}{dim 6} & $D_\mu \overline{Q_L}^i \gamma^\mu \psi^j H^\dagger_iH_j$ \ , \ $ \overline{Q_L}^i \gamma^\mu D_\mu \psi^j H^\dagger_iH_j$ & \multirow{4}{*}{($3,2,1/6$)}\tabularnewline
 & $\overline{Q_L}^i \gamma^\mu \psi^j H^\dagger_iD_\mu H_j$ & \tabularnewline
  & $D_\mu \overline{Q_L}^i \gamma^\mu \psi_i H^{\dagger j} H_j$ \ , \ $ \overline{Q_L}^i \gamma^\mu D_\mu \psi_i H^{\dagger j} H_j$ & \tabularnewline
  & $\overline{Q_L}^i \gamma^\mu \psi_i H^{\dagger j} D_\mu H_j$ & \tabularnewline
 \hline

\multirow{2}{*}{dim 6} & $D_\mu \overline{Q_L}^i \gamma^\mu \psi_{ijk} H^{\dagger j}H^{\dagger k}$ & \multirow{2}{*}{($3,4,7/6$)}\tabularnewline
 & $\overline{Q_L}^i \gamma^\mu \psi_{ijk} H^{\dagger j}D_\mu H^{\dagger k}$ & \tabularnewline
\hline
\multirow{2}{*}{dim 6} & $D_\mu \overline{Q_L}^i \gamma^\mu \psi_j H^\dagger_iH^{\dagger j}$ & \multirow{2}{*}{($3,2,7/6$)}\tabularnewline
& $\overline{Q_L}^i \gamma^\mu \psi_j H^\dagger_i D_\mu H^{\dagger j}$ & \tabularnewline
\hline

dim 6 & $\overline{u} \gamma^\mu \psi^{ij} H_iD_\mu H_j$ & ($3,3,-1/3$)\tabularnewline
\hline
dim 6 & $\overline{d} \gamma^\mu \psi^{ij} H_iD_\mu H_j$ & ($3,3,-4/3$)\tabularnewline
\hline
dim 6 & $\overline{u} \gamma^\mu \psi^{j}_i H^{\dagger i}D_\mu H_j$ \ , \  $\overline{u} \gamma^\mu D_\mu \psi^{j}_i H^{\dagger i}H_j$ & ($3,3,2/3$)\tabularnewline
\hline
dim 6 & $\overline{d} \gamma^\mu \psi^{j}_i H^{\dagger i}D_\mu H_j$ \ , \ $\overline{d} \gamma^\mu D_\mu \psi^{j}_i H^{\dagger i} H_j$ & ($3,3,-1/3$)\tabularnewline
\hline
dim 6 & $\overline{u} \gamma^\mu \psi H^{\dagger i}D_\mu H_i$ & ($3,1,2/3$)\tabularnewline
\hline
dim 6 & $\overline{d} \gamma^\mu \psi H^{\dagger i}D_\mu H_i$ & ($3,1,-1/3$)\tabularnewline
\hline
dim 6 & $\overline{u} \gamma^\mu \psi_{ij} H^{\dagger i} D_\mu H^{\dagger j}$ & ($3,3,5/3$)\tabularnewline
\hline
dim 6 & $\overline{d} \gamma^\mu \psi_{ij} H^{\dagger i} D_\mu H^{\dagger j}$ & ($3,3,2/3$)\tabularnewline
\hline
\end{tabular}
\\
Table 13. Operators containing $DQ \psi HH$
\end{center}

\subsubsection{Operators with $\boldsymbol{DDQ\psi H}$}
The next dimension 6 operator that we will examine contains one quark, a LEX fermion, one Higgs field, and two covariant derivatives. Again, the quark can be left- or right-handed. Also repeated from previous arguments, we can choose to work with only one two-derivative Lorentz structure. The other two are redundant with each other and the operator form $Q\psi HW$, which was discussed earlier in the paper. Here, we again choose to work with the $\slashed{D}\slashed{D}$ Lorentz structure. Again, integration by parts reduces six possible operators to three independent ones. We choose to keep
\begin{equation}
    \slashed{D}\slashed{D} Q \psi H, \quad \slashed{D}Q \slashed{D}\psi H, \quad Q \slashed{D}\slashed{D}\psi  H \ .
\end{equation}

Following from previous arguments, the first of these operators is redundant with those of type $DQ\psi HH$ due to the SM fermion equations of motion. For the other two operators, we remember the general fermion equation of motion.
\begin{equation}
    \slashed{D}\psi = m\psi
\end{equation}
To look at this applied, we note that $Q \slashed{D}\slashed{D}\psi  H = m Q \slashed{D}\psi H$. Similarly, there are two options for $\slashed{D}Q \slashed{D}\psi H$. This operator is equivalent to either $(y_d \overline{d_R} H + y_u \overline{u_R} H^\dagger)\slashed{D}\psi H$ or$\slashed{D}Q m\psi H$.  As such, all operators of type $DDQ\psi H$ reduce to those of type $DQ\psi HH$ or $DQ\psi H$. This operator type is fully redundant.

\subsubsection{Operators with $\boldsymbol{{Q} D D D\psi}$}

Finally, we examine dimension 6 operators with one SM quark, a LEX fermion, and three covariant derivatives. For operators of this type, the quark must be left-handed. There are three categories of Lorentz structures for operators of this type.
The first category contains operators where two of the derivatives have their Lorentz indices contracted with each other, and the third derivative has its index contracted with a gamma matrix.
The second and third categories involve three derivatives with no shared Lorentz indices. The second category that we will discuss contracts the derivatives with three gamma matrices, while the third category contracts the derivatives with a $\sigma^{\mu\nu}$ and a single gamma matrix.

Starting with the first category listed above, we must consider all possible ordering of the derivatives and gamma matrices, along with all options of how many derivatives act on each field. The options are shown below. We note that, due to integration by parts, all operators with derivatives acting on more than one field are directly related to those where all derivatives act on the same field.
\begin{align}
    \begin{split}
        D_\mu D_\alpha D^\alpha \overline{Q} \gamma^\mu\psi = D_\alpha D^\alpha \overline{Q} \gamma^\mu D_\mu\psi &= D^\alpha \overline{Q}\gamma^\mu D_\alpha D_\mu \psi = \overline{Q} \gamma^\mu D_\alpha D^\alpha D_\mu\psi 
        \\
         D_\alpha D_\mu D^\alpha \overline{Q} \gamma^\mu \psi = D_\mu  D^\alpha \overline{Q} \gamma^\mu D_\alpha \psi &= D^\alpha \overline{Q} \gamma^\mu D_\mu D_\alpha \psi = \overline{Q} \gamma^\mu D^\alpha D_\mu D_\alpha \psi
         \\
         D_\alpha D^\alpha D_\mu \overline{Q} \gamma^\mu\psi = D^\alpha D_\mu \overline{Q} \gamma^\mu D_\alpha \psi &= D_\mu \overline{Q} \gamma^\mu D^\alpha D_\alpha \psi = \overline{Q} \gamma^\mu D_\mu D^\alpha D_\alpha \psi
    \end{split}
\end{align}
We choose to work with the first option in each line of Eq. 25, and will show that these are all redundant with each other and other operators in the portal. We first show that we can get rid of two of the operators, and then do more extensive calculations to drop the final operator.

Using the anti-commutation relations for covariant derivatives, the first operators on each line of Eq. 25 are related to each other:
\begin{align}
    \begin{split}
        D_\mu D_\alpha D^\alpha \overline{Q} \gamma^\mu\psi &= D_\alpha D_\mu D^\alpha \overline{Q} \gamma^\mu\psi + \frac{1}{ig}F_{\mu\alpha}D^\alpha \overline{Q} \gamma^\mu\psi
        \\
        D_\alpha D_\mu D^\alpha \overline{Q} \gamma^\mu\psi &= D_\alpha D^\alpha D_\mu \overline{Q} \gamma^\mu\psi + D_\alpha (\frac{1}{ig}F_{\mu\alpha}\overline{Q}) \gamma^\mu\psi
    \end{split}
\end{align}
The last terms in both lines of Eq. 26 are of type $QD\psi F$, which is found in Section 5.1 of this paper. As such, we have two equations with three "new" operator types. This allows us to discard two of these, and continue working with only one. Here, we choose to continue to examine $D_\alpha D^\alpha D_\mu \overline{Q} \gamma^\mu\psi = D^2 \overline{Q} \overleftarrow{\slashed{D}}\psi$.

We can utilize the quark equations of motion to simplify this operator into those of a form already discussed in this paper. Specifically, we have
\begin{align}
    \begin{split}
        D^2 \overline{Q} \overleftarrow{\slashed{D}}\psi = &D^2 (y_d \overline{d}H + y_u \overline{u}H^\dagger)\psi
        \\ 
        =&(D^2 \overline{d} + 2D^\alpha\overline{d}D_\alpha H + \overline{d}D^2 H  \\ & + D^2 \overline{u} + 2D^\alpha\overline{u}D_\alpha H^\dagger + \overline{u}D^2 H^\dagger)\psi \ .
    \end{split}
\end{align}
In Eq. 27, the terms $D^2\overline{d}\psi$ and $D^2\overline{u}\psi$ do not contribute to the W-quark portal. All other terms in this expression have already been considered under operators of type $DDqH\psi$, which were discussed in Section 5.4.1. There are thus no new operators with this structure.

Next, we examine operators with three derivatives and three gamma matrices. There are six operator options with this structure. Specifically, we have
\begin{align}
    \begin{split}
        \textcircled{1} &= D_\mu D_\nu D_\delta \overline{Q} \gamma^\delta\gamma^\nu\gamma^\mu\psi \quad 
        \textcircled{2} =D_\mu D_\delta D_\nu \overline{Q} \gamma^\delta\gamma^\nu\gamma^\mu\psi  
        \\
        \textcircled{3} &= D_\mu D_\nu D_\delta \overline{Q} \gamma^\delta\gamma^\nu\gamma^\mu\psi \quad
        \textcircled{4} = D_\nu D_\delta D_\mu \overline{Q} \gamma^\delta\gamma^\nu\gamma^\mu\psi
        \\
        \textcircled{5} &=D_\delta D_\mu D_\nu \overline{Q} \gamma^\delta\gamma^\nu\gamma^\mu\psi \quad 
        \textcircled{6} = D_\delta D_\nu D_\mu \overline{Q} \gamma^\delta\gamma^\nu\gamma^\mu\psi \ .
    \end{split}
\end{align}
Similarly to the case above, all other operators of this structure with one, two, or three derivatives acting on $\psi$ are directly equivalent to the operators listed in Eq. 28 through integration by parts. 

All of the operators in Eq. 28 are also redundant. We will show this, starting with the operator marked with $\textcircled{1}$. This can be done simply by using the equations of motion and integration by parts:
\begin{align}
    \begin{split}
        \textcircled{1} &= D_\mu D_\nu \overline{Q} \overleftarrow{\slashed{D}}\gamma^\nu\gamma^\mu\psi  = D_\nu (y_d \overline{d}H + y_u \overline{u}H^\dagger) \gamma^\nu\gamma^\mu D_\mu\psi 
        \\
        &= (y_d \overline{d}H + y_u \overline{u}H^\dagger) \gamma^\nu D_\nu \gamma^\mu D_\mu\psi = (y_d \overline{d}H + y_u \overline{u}H^\dagger) \slashed{D}\slashed{D}\psi \ .
    \end{split}
\end{align}
This operator is thus redundant with those of type $DDqH\psi$ in Section 5.4.1.

The operator labeled $\textcircled{3}$ can also have the quark equation of motion applied. After this, the covariant derivative commutation relations can be applied to the remaining derivatives. 
This results in 
\begin{align}
\begin{split}
\textcircled{3} &= D_\mu D_\nu (y_d \overline{d}H + y_u \overline{u}H^\dagger) \gamma^\nu\gamma^\mu \psi + \frac{1}{ig}F_{\nu\mu} (y_d \overline{d}H + y_u \overline{u}H^\dagger)\gamma^\nu\gamma^\mu\psi 
\\
&= \textcircled{1} + \frac{1}{ig}F_{\nu\mu} (y_d \overline{d}H + y_u \overline{u}H^\dagger)\frac{1}{2}\sigma^{\nu\mu}\psi \ .
\end{split}
\end{align}
In the last equality, we used the antisymmetric properties of $F_{\nu\mu}$ to pull out the antisymmetric part of $\gamma^\nu\gamma^\mu$. 
The operator $\textcircled{3}$ is thus redundant with the operator labeled $\textcircled{1}$ from Eq. 28 and operators of the type $q\psi WH$ from Section 3.2. It can be removed from the portal.

Operator $\textcircled{2}$ from Eq. 28 can be examined next. This operator can be integrated by parts, and then the fermion $\psi$ equations of motion may be applied. Further, using the derivative commutator and $\overline{Q}$ equations of motion gives us the result
\begin{equation}
    \textcircled{2} = m(y_d\overline{d}D_\nu H + y_u\overline{u}D_\nu H^\dagger)\gamma^\nu\psi + \frac{1}{2ig}F_{\delta\nu} \overline{Q}\sigma^{\delta\nu}\psi \ .
\end{equation}
We have also removed terms from Eq. 31 which do not contribute to the W-quark portal. The first two remaining terms are redundant with the operators of type $Dq\psi H$ from Section 5.3, while the third term is redundant with those of type $FQ\psi$ from Section 3.1.

Operators $\textcircled{4}$ through $\textcircled{6}$ from Eq. 24 can also be shown to be redundant through similar methods. Additionally, the relation $\{\gamma^\mu,\gamma^\nu\} = 2g^{\mu\nu}$ is also useful here. It can be shown that
\begin{align}
    \begin{split}
        \textcircled{4} &= \textcircled{3} + \frac{2}{ig}F_{\delta\mu}\overline{Q}\gamma^\delta D^\mu \psi - \frac{m}{2ig}F_{\delta\mu} \overline{Q}\sigma^{\mu\nu}\psi \ ,
        \\
        \textcircled{5} &= \textcircled{2} + \frac{2}{ig}F_{\delta\mu} D^\delta \overline{Q}\gamma^\mu \psi - \frac{1}{2ig} F_{\delta\mu } (y_d \overline{d}H + y_u \overline{u}H^\dagger) \sigma^{\delta\mu}\psi \ ,
        \\
        \textcircled{6} &= \textcircled{4} + \frac{1}{ig}F_{\delta\nu}(y_d \overline{d}H + y_u \overline{u}H^\dagger)\sigma^{\delta\nu}\psi - \frac{4}{ig}F_{\delta\nu }D^\delta \overline{Q}\gamma^\nu\psi  \ .    
    \end{split}
\end{align}
Operator $\textcircled{4}$ is thus redundant with operator $\textcircled{3}$ and those of types $DQ\psi F$ and $FQ\psi$, from Sections 5.1.1 and 3.1, respectively. Operator $\textcircled{5}$ is redundant with operator $\textcircled{2}$ and those of types $DQ\psi F$ and $q\psi WH$. These operators are from Sections 5.1.1 and 3.1. Last, operator $\textcircled{6}$ is thus redundant with operator $\textcircled{4}$ and those from Sections 3.1 and 5.1.1. 

Finally, we can address operators with three derivatives, a $\sigma^{\mu\nu}$, and a gamma matrix. Already taking into consideration the antisymmetric nature of $\sigma$, there are also six potential operators:
\begin{align}
    \begin{split}
        &D_\mu D_\nu D_\delta \overline{Q}\sigma^{\mu\nu } \gamma^\delta \psi , \quad  
        D_\mu D_\delta D_\nu \overline{Q}\sigma^{\mu\nu } \gamma^\delta \psi , \quad 
        D_\delta D_\mu D_\nu \overline{Q}\sigma^{\mu\nu } \gamma^\delta \psi , 
        \\
        &D_\mu D_\nu D_\delta \overline{Q}\gamma^\delta \sigma^{\mu\nu }  \psi , \quad  
        D_\mu D_\delta D_\nu \overline{Q}\gamma^\delta  \sigma^{\mu\nu } \psi , \quad 
        D_\delta D_\mu D_\nu \overline{Q}\gamma^\delta \sigma^{\mu\nu }  \psi . 
    \end{split}
\end{align}
However, due to the definition of $\sigma^{\mu\nu}$, each of these operators is equal to the sum of two of the operators with three gamma matrices. For example,
\begin{equation}
    D_\mu D_\nu D_\delta \overline{Q}\sigma^{\mu\nu } \gamma^\delta \psi = D_\mu D_\nu D_\delta \overline{Q}(\gamma^\mu\gamma^\nu - \gamma^\nu\gamma^\mu) \gamma^\delta \psi = \textcircled{1}- \textcircled{3} \ .
\end{equation}
These operators can thus be removed from the portal as well. There are no new operators of type $QDDD\psi$.

\subsection{Some Collider Phenomenology with Fermionic LEX Fields and Derivatives}
The operators including both fermionic LEX states and covariant derivatives are the first where we see couplings to gluons. It is also in these operators including gluon couplings where we find an operator in the W-quark portal that reaches an SU(3) charge of 15. This operator, $\frac{1}{\Lambda^2}\overline{Q_L}^i\gamma^\mu \overleftarrow{D^\nu} \psi_i G_{\mu \nu}$, could be very interesting to study due to the Clebsh Gordan coefficients that accompany a 15-plet of SU(3). 

We also find operators which lead to interaction vertices that contain two Higgs bosons. While either or both of the Higgs could be set to their vev, this is not required and could lead to interesting vertices. Both of these new types of interaction vertices are shown in Fig. 9.
\begin{center}
    \includegraphics[scale=.1]{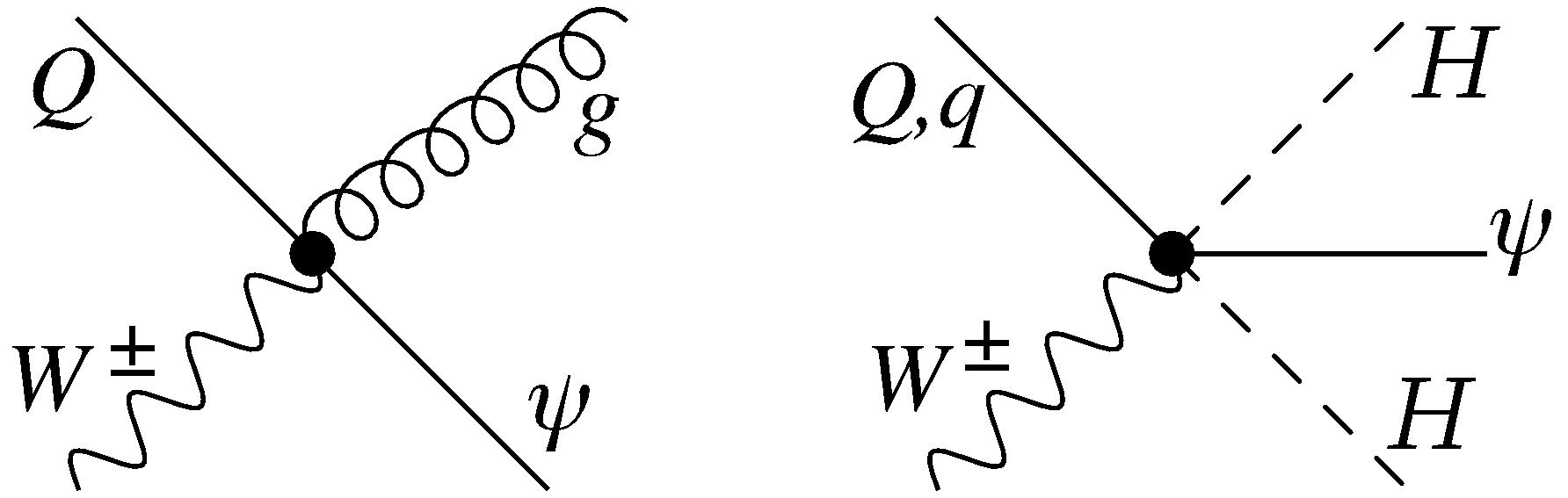} \\
    Figure 9. Some interaction vertices due to the fermionic LEX operators that include derivatives
\end{center}

Also within this section, the operators of type $QD\psi$ are interesting, as they are the lowest-dimensional operators within this portal. In this operator, $\psi$ has the same quantum numbers as a left-handed quark: ($3,2,1/6$) While there are several other operator types that lead to the three-particle vertex with only a W boson, quark, and LEX fermion, it is the operators of this type that will likely result in processes with the largest cross sections, and thus discovery potential, as less powers of the effective cutoff are required.

\section{LHC Exploration}
We will now look at an example of the interesting nature of the models that the W-quark portal provides access to. Specifically, we here analyze a bit of LHC phenomenology for  one specific model. We have chosen a simple but interesting example of an SU(3) charged LEX state in a higher dimensional representation of SU(2), the fermionic SU(3) triplet SU(2) quadruplet with SM representation ($3,4, 1/6$). The specific dimension 5 operator we analyze is given by
\begin{equation}
\frac{1}{\Lambda}\overline{Q}^k\sigma^{\mu \nu} \psi_{i j k} W_{\mu \nu}^{i j}  \ .
\end{equation}
In Eq. 35, the Roman letters indicate fundamental SU(2) indices, while the Greek letters indicate Lorentz indices. The LEX state is $\psi$. The quadruplet contains four components with charges ($5/3, 2/3, -1/3, -4/3$). The field strength tensor contains the SU(2) gauge fields. After electro-weak symmetry breaking, this operator may be expanded into various charge components. This yields the Lagrangian terms
\begin{align}
\frac{g}{6\text{s}_\text{w}} \Bigl[  e& \Bigl( 2\sqrt{3} \overline{\psi}_{2 } d  W_\mu W^\dagger_\nu 
 \\
\nonumber & +  \sqrt{2}\text{c}_\text{w} W^\dagger_\mu Z_\nu \Big\{\left( 3\overline{\psi}_{1} d + \sqrt{3} \overline{\psi}_{2} u\right)  
 - \left( \sqrt{3}\overline{d} \psi_{3} + 3\overline{u}\psi_{4} \right) 
 \\
 \nonumber & + \sqrt{2} \text{s}_\text{w} A_\mu \Big\{   -\left[\left(\sqrt{3} \overline{\psi}_{3} d + 3 \overline{\psi}_{4} u\right) W_\nu + \left(3\overline{\psi}_1 d + \sqrt{3} \overline{\psi}_2 u \right) W^\dagger_\nu \right] 
 \\
 \nonumber & \qquad \qquad \quad +  \left[ \left(3 \overline{d} \psi_{1} + \sqrt{3} \overline{u} \psi_{2} \right)W_\nu + \left(\sqrt{3}\overline{d} \psi_{3} + 3\overline{u} \psi_{4} \right) W^\dagger_\nu\right] 
 \\
 \nonumber & + W_\mu \Big\{2\sqrt{3} \overline{\psi}_{3} u  W^\dagger_\nu  + \sqrt{2} \text{c}_\text{w} \left( \sqrt3 \overline{\psi}_{3} d + 3\overline{\psi}_{4} u \right)  Z_\nu 
 \\
 \nonumber & \qquad +  \left[2\sqrt{3} \left(\overline{d} \psi_{2} + \overline{u} \psi_{3}\right) W^\dagger_\nu - \sqrt{2} \text{c}_\text{w} \left(3 \overline{d} \psi_{1} + \sqrt{3}\overline{u} \psi_{2} \right) Z_\nu  \right] 
 \Big\}   \Big) 
\\
 \nonumber + & i \sqrt{2} \text{s}_\text{w}\partial_\nu W_\mu 
 \left( \left[ \sqrt{3}\overline{\psi}_{3} d + 3 \overline{\psi}_{4} u \right]  
 -  \left[ 3 \overline{d} \psi_{1} + \sqrt{3} \overline{u} \psi_{2} \right] 
 \right)
 \\
\nonumber - & 2 i \sqrt{3} \left(\text{s}_\text{w}^2\partial_\nu A_\mu + \text{c}_\text{w}\text{s}_\text{w} \partial_\nu Z_\mu \right)
 \left( \left[ \overline{\psi}_{2} d +\overline{\psi}_{3} u \right]  
 + \left[ \overline{d} \psi_{2} + \overline{u} \psi_{3} \right] \right.
 \\ \nonumber
 -&i\sqrt{2} \text{s}_\text{w}\partial_\nu W^\dagger_\mu 
 \left(
 \left[ 3\overline{\psi}_{1} d + \sqrt{3}\overline{\psi}_{2} u \right]  
 -  \left[ \sqrt{3} \overline{d} \psi_{3} + 3\overline{u} \psi_{4} \right] 
 \right)
 \Bigl. \Bigr] \left[\gamma^\mu,\gamma^\nu\right]
\end{align}
where the lowest charge component is denoted by $\psi_1$, and the highest by $\psi_4$.

From these interaction terms we can find several possible LHC production processes for the quadruplet components.  While we note that pair production is always allowed by gluon fusion, there are several channels for single quadruplet production. The last three lines of Eq. 36 give single gauge boson interactions- these terms contain one electroweak gauge boson, one quark, and a LEX quadruplet component. 
We denote the highest charge component of the multiplet, with electric charge $5/3$, as $\psi_4$. This state couples to a negative W boson and anti-up quark  ($W^- \overline{u}$). The state $\psi_3$ with charge $2/3$ couples to $W^{-} \overline{d} $. Meanwhile, the state $\psi_2$ with charge $-1/3$ couples to  $W^{+} \overline{u} $ and the state  $\psi_1$ with charge $-4/3$ couples to  $W^{+} \overline{d} $.  These three point vertices between the LEX quadruplet, charged W boson, and quarks are shown in the upper left corner of Fig. 5.  LHC production follows through s-channel and t-channel processes as shown in Fig. 6 and Fig. 7, respectively.  The s-channel process involves quark fusion into an off-shell W boson, which  then produces a LEX-quark pair. The t-channel process involves quark fusion with a t-channel exchange of a charged W boson, producing a final state quadruplet and quark. These processes give $\overline{q}q^{'}\rightarrow \psi \overline{q}/ q \overline{\psi}$. 

We also have an available process of $\psi$ vector boson-associated production. This process arrives from quark-gluon fusion into an off-shell quark, 
$qg\rightarrow q^{*}\rightarrow \psi + V$. In Fig. 10, we compare various LHC pair and single production cross-sections for the quadruplet state. 
\begin{center}
\includegraphics[scale=.5]{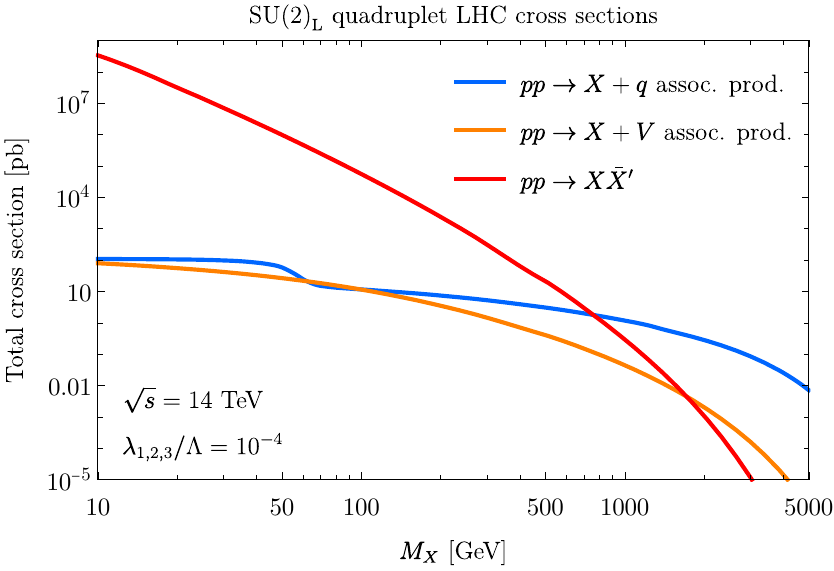} \\
Figure 10. Cross sections for single and pair production of quadruplet states at the 14 TeV LHC
\end{center}

In Fig. 10, $X$ denotes all components of the LEX quadruplet. To create this graph, we have generated a UFO \cite{UFO}for our model using \textsc{FeynRules} \cite{FR_OG, FR_2}. We implemented this UFO within \textsc{MadGraph5\texttt{\textunderscore}aMC@NLO} (\textsc{MG5\texttt{\textunderscore}aMC}) \cite{Alwall:2014hca} to compute the leading order cross section for all quadruplet $\psi$ states at the 14 TeV LHC. We note a few interesting features from this graph. First, though the quadruplet pair production cross section dominates all other production modes at low masses, we see that higher mass states ---above 1--1.5 TeV--- have higher production cross sections in single production modes. At higher LEX masses, there is a large kinematic suppression of pair produced events, while there is still adequate phase-space for the light mass quadruplet-quark pair.  

Next, we see that throughout the mass range shown, associated production of quadruplet-quark final states outperforms associated production of quadruplet-gauge boson events. Naively, one might expect quark-gluon fusion events to carry a higher cross section than the quark-quark fusion processes that result in $\psi q$ associated production. However, this is not the case for high mass $\psi$.  At the LHC, the quarks dominate the parton distribution function at high momentum fraction. As such, these energetic  quarks dominate the quadruplet production cross sections. As an illustration of this, we loaded our event files generated by \textsc{MG5\texttt{\textunderscore}aMC} into $\textsc{MadAnalysis5}$ \cite{Conte_2013,Conte_2014,Conte_2018} to examine some kinematic distributions. Specifically, we have plotted the momentum distributions for the incoming and outgoing quarks in the quadruplet-quark associated production process. These distributions can be found in Fig. 11.

\begin{center}
\includegraphics[scale=.33]{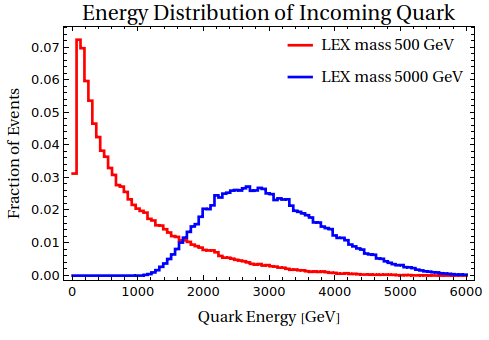} ~~\includegraphics[scale=.33]{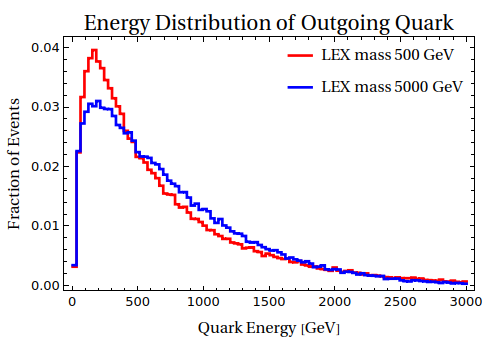} \\
Fig 11. Incoming and outgoing quark energies for $qq\rightarrow \psi q$ events at the 14 TeV LHC
\end{center}

In order to produce our events, we require quarks with incoming energy roughly $m_{\psi}/2$. These high momentum quark partons dominate the pdf's compared to high energy gluons. Thus in the $q+\psi$ channel we still see production cross-sections above 1 fb, even for quadruplet masses up to 5 TeV. We also note that the outgoing quark in the event can also be hard, as shown in the right panel of Fig. 11. We find a significant fraction of events with an outgoing quark energy above 500 GeV. These hard outgoing quarks can be an important discriminating feature in a dedicated LHC search for associated production of the quadruplet.

Through our operator, the components of the quadruplet always have an open decay channel available into a quark and electroweak gauge boson, $\psi\rightarrow q+V$. The effective operator allows interesting stand out decays that involve hard photons, missing energy or leptons. For example, $\psi_2\rightarrow d \gamma/Z$ and $\psi_3\rightarrow u \gamma/Z$ are allowed, with possible $Z\rightarrow \nu\nu, \ell\ell$ channels. In addition, decays with W's, for example  $\psi_3\rightarrow d W^{+}$, may also involve leptons. 

We note that if the masses of the charge components of the quadruplet  can be split, then there are additional decay channels opened between the charge components due to the tree level electroweak coupling. One potential process that could be allowed by mass splitting would be $\psi_3\rightarrow W^+ \psi_2$. This opens the possibility for multi-body cascade decays of the quadruplet states, eg $\psi_3\rightarrow W^+ \psi_2\rightarrow W^+ W^+ \psi_1$. This mass splitting, however, must be achieved by coupling the quadruplet to a state involved in electro-weak symmetry breaking. The simplest possibility involves coupling the quadruplets to Higgs fields at dimension 5: $\frac{1}{\Lambda}H\psi H^{\dagger}\overline{\psi}$. There are several possibilities for the contraction of SU(2) indices here that will result in splittings between different components of the quadruplet multiplet. The study of the detailed phenomenology of this model is a potential topic for further research.

\section{Conclusion}
In this paper, we presented the W-quark portal to BSM physics. We cataloged all new BSM states that appear in interactions up to dimension 6 that involve at least one quark and one gluon. We have shown that this portal provides access to exotic states such as exotic scalar SU(3) sextets and octets, scalar SU(3) triplets including those of higher hypercharge, fermionic SU(3) sextets and 15-plets, and states of higher SU(2) charge including quadruplets and quintuplets of SU(5). In addition, we have cataloged effective interactions for particles in various BSM paradigms such as exotics Higgses, heavy vector-like quarks, Manohar-Wise octets, and more. Many of the new interactions listed contain interesting features such as sets of leptonic operators that may lead to processes with one or multiple hard leptons. 

This catalog of effective interactions may yield new phenomenological searches for particles in these paradigms. We have shown how we have removed redundant operators from our catalog through integration by parts procedures and field theory identities, and we have justified our use of phenomenological operator basis to make the most general definition of the W-quark portal. We have outlined some important collider production processes for the LHC and muon colliders for several types of exotic states. These include new s-channel and t-channel quark fusion processes that involve W boson exchanges. 

As an example of how interesting and intricate new W-q portal models are, we have performed a brief phenomenological analysis of a new accessible particle in a higher dimensional representation of SU(2). Specifically, we analyzed a model that contains an exotic fermionic SU(3) triplet that is a quadruplet under SU(2). We have written the Lagrangian for this process and computed LHC production cross sections for the various charged states in the multiplet. We focus our subsequent discussion on our new quark fusion channel in which a single exotic particle is produced in association with a hard SM quark.

There are many opportunities to build on this work. The first is to do some extended phenomenological explorations of the most interesting exotic states which are reachable through the W-quark portal. Specifically, it would be interesting to focus on the states in higher dimensional representations of SU(3) and/or SU(2). An immediate candidate would be to continue the phenomenological studies of the SU(2) quadruplet fermion that we have begun in this paper. Other interesting states to study are those with double baryon number, those states in the 15-plet representation of SU(3), and the states in representations higher than triplet in SU(2). Another interesting avenue for study is to look at the general Lagrangians and build UV completions of some of the under-explored (or as yet unexplored) exotic particles in this catalog. One recent model the authors have examined is the scalar bi-adjoint (adjoint of both SU(3) and SU(2) gauge groups); we look forward to model building with different exotic states. Studies of precision constraints such as flavor and electro-weak precision for general charges of exotics would also be interesting. Some  work has been done in this direction \cite{Carpenter:2022oyg,Fortes:2013dba}, but there is much more ground to cover. Another possibility for study would be to explore complete EFT operator catalogs for any of the exotic particles that reach beyond the W-quark portal operators.  We find that the phenomenological landscape of exotic EFTs is still wide open for exploration. 

\bibliographystyle{JHEP}

\bibliography{biblq.bib}

\end{document}